\documentclass[acmsmall,nonacm]{acmart}

\pdfoutput=1

\usepackage{adjustbox}
\usepackage{graphicx}
\usepackage{microtype}
\usepackage{multirow}
\usepackage[all]{nowidow}
\usepackage{tabularx}

\usepackage{titlesec}
\titlespacing{\subparagraph}{0pt}{.5\baselineskip}{1ex}
\titleformat{\subparagraph}[runin]{\bfseries}{\thesubparagraph}{0pt}{}[:]

\usepackage{xspace}
\newcommand{\ie}{\textit{i.e.,}\xspace}
\newcommand{\eg}{\textit{e.g.,}\xspace}  

\newcommand*{\ifuselandscape}[2]{#2}

\ifuselandscape{\usepackage{pdflscape}}{\usepackage{rotating}}

\makeatletter
\newcommand*{\col}[3]{%
	\edef\@propCounter{#1Counter}%
	\ifcsname c@#1Counter\endcsname \else \newcounter{\@propCounter} \fi
	\refstepcounter{\@propCounter}\def\@currentlabel{#1\arabic{\@propCounter}}\label{col:#2}%\
	\def\@currentlabel{\textit{#3}}\label{col-text:#2}%
	\def\@currentlabel{\ref{col:#2}:~\ref{col-text:#2}}\label{col-full:#2}%
}
\makeatother

\newcommand*{\setuptable}{
	\renewcommand{\arraystretch}{1}
	\setlength\tabcolsep{0em}
	\small
	\centering
}

\newlength{\cellsize}
\newcommand*{\cell}[1]{\parbox[m][\cellsize][c]{\cellsize}{\centering#1}}
\newcommand*{\imagecell}[2][.9]{\cell{\includegraphics[width=#1\cellsize,height=#1\cellsize]{#2}}}

\newcommand*{\head}[2]{\multicolumn{1}{#1}{#2}}
\newcommand*{\headtilt}[1]{\rotatebox{90}{#1}}

\newcolumntype{L}{>{\hspace{.25em}}l<{\hspace{.25em}}}
\newcolumntype{R}{>{\hspace{.25em}}l<{\hspace{.25em}}}
\newcolumntype{C}{>{\centering\arraybackslash}p{\cellsize}}

\usepackage{wasysym}
\newcommand*{\fullsymbol}{\CIRCLE}
\newcommand*{\prtsymbol}{\LEFTcircle}
\newcommand*{\nonesymbol}{\Circle}
\newcommand*{\nasymbol}{}

\newcommand*{\full}{\cell{\fullsymbol}}
\newcommand*{\prt}{\cell{\prtsymbol}}
\newcommand*{\none}{\cell{\nonesymbol}}
\newcommand*{\na}{\cell{\nasymbol}}

\newcommand*{\high}{\imagecell[1]{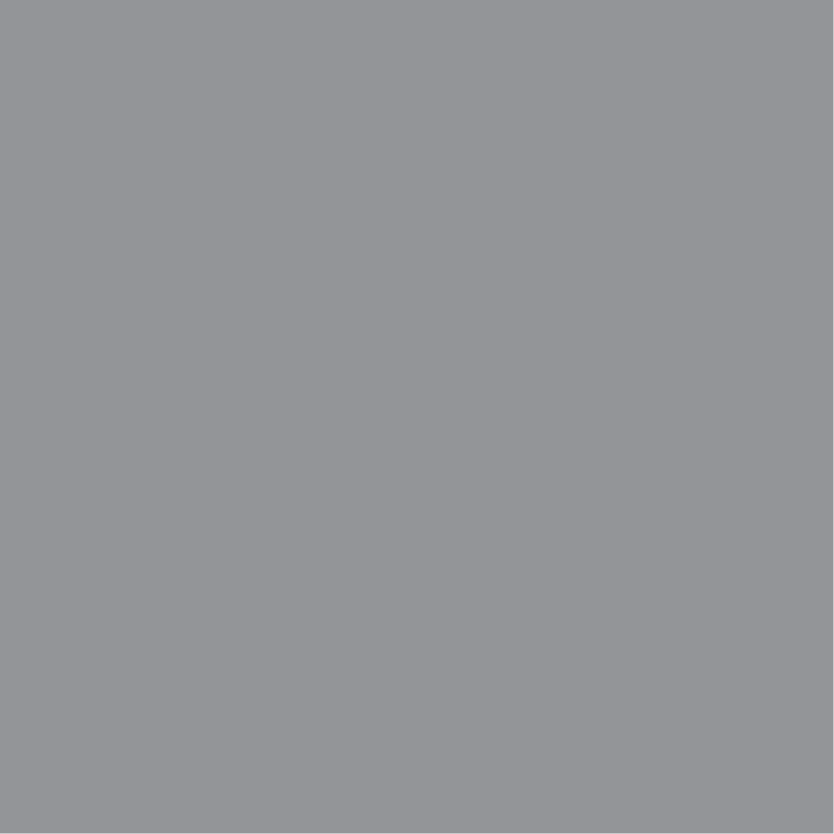}}
\newcommand*{\highprt}{\imagecell{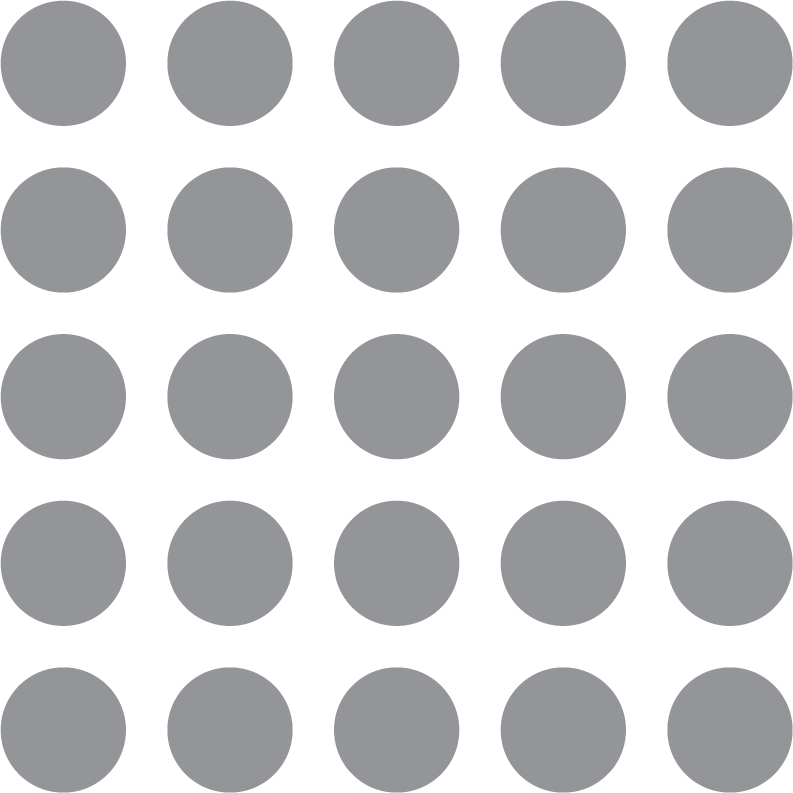}}
\newcommand*{\low}{\imagecell{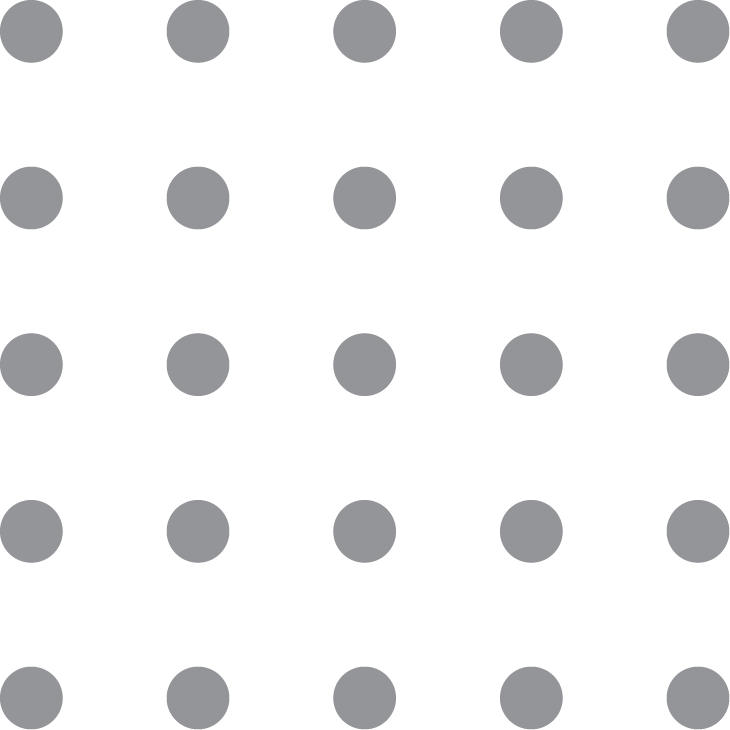}}
\newcommand*{\anti}{\imagecell{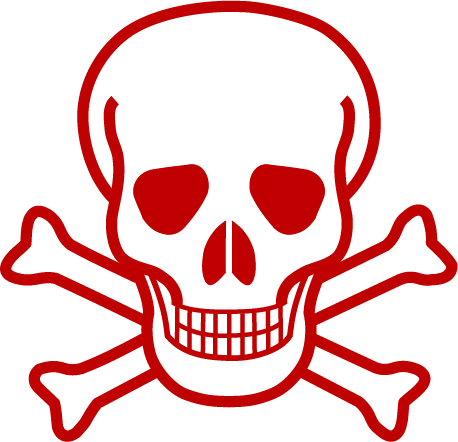}}
\newcommand*{\tussle}{\imagecell{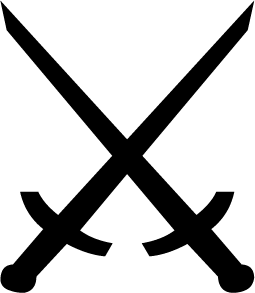}}

\newcommand*{\ratingtext}[1]{\parbox[m][1em][c]{1em}{\centering\includegraphics[width=.9em,height=.9em]{#1}}}
\newcommand*{\hightext}{\ratingtext{cells/high-priority.png}}
\newcommand*{\highprttext}{\ratingtext{cells/high-priority-prt.png}}

\newcommand*{\antitext}{\ratingtext{cells/skull-red.png}}
\newcommand*{\tussletext}{\ratingtext{cells/swords.png}}

\setcopyright{rightsretained}
\copyrightyear{2021}
\acmYear{2021}
\acmDOI{}

\acmJournal{CSUR}
\acmVolume{}
\acmNumber{}
\acmArticle{}
\acmMonth{10}

\begin{document}

\title[SoK: Securing Email---A Stakeholder-Based Analysis (Extended Version)]{SoK: Securing Email---A Stakeholder-Based Analysis\\(Extended Version)\textsuperscript{\textasteriskcentered}}
%\title{A Survey of Secure Email Systems and Stakeholders}

\author{Jeremy Clark}
\email{j.clark@concordia.ca}
\affiliation{%
	\institution{Concordia University}
	\country{Canada}
	\state{Quebec}
	\city{Montreal}
}

\author{P.C. van Oorschot}
\email{paulv@scs.carleton.ca}
\affiliation{%
	\institution{Carleton University}
	\country{Canada}
	\state{Ontario}
	\city{Ottawa}
}

\author{Scott Ruoti}
\email{ruoti@utk.edu}
\affiliation{%
	\institution{University of Tennessee}
	\country{United States of America}
	\state{Tennessee}
	\city{Knoxville}
}

\author{Kent Seamons}
\email{seamons@cs.byu.edu}
\affiliation{%
	\institution{Brigham Young University}
	\country{United States of America}
	\state{Utah}
	\city{Provo}
}

\author{Daniel Zappala}
\email{zappala@cs.byu.edu}
\affiliation{%
	\institution{Brigham Young University}
	\country{United States of America}
	\state{Utah}
	\city{Provo}
}

%\renewcommand{\shortauthors}{Clark et al.}

% = = = = = Abstract = = = = = %

% !TEX root = ../main.tex

\begin{abstract}
While email is the most ubiquitous and interoperable form of online communication today, it was not conceived with strong security guarantees, and the ensuing security enhancements are, by contrast, lacking in both ubiquity and interoperability.
This situation motivates our research.
We begin by identifying a variety of stakeholders who have an interest in the current email system and in efforts to provide secure solutions.
We then use the tussle among stakeholders to explain the evolution of fragmented secure email solutions undertaken by industry, academia, and independent developers.
We also evaluate the building blocks of secure email---cryptographic primitives, key management schemes, and system designs---to identify their support for stakeholder properties.
From our analysis, we conclude that a one-size-fits-all solution is unlikely.
Furthermore, we highlight that vulnerable users are not well served by current solutions, account for the failure of PGP, and argue that secure messaging, while complementary, is not a fully substitutable technology.
\end{abstract}

\begin{CCSXML}
	<ccs2012>
	<concept>
	<concept_id>10002951.10003260.10003282.10003286.10003287</concept_id>
	<concept_desc>Information systems~Email</concept_desc>
	<concept_significance>500</concept_significance>
	</concept>
	<concept>
	<concept_id>10002978.10003014.10003015</concept_id>
	<concept_desc>Security and privacy~Security protocols</concept_desc>
	<concept_significance>500</concept_significance>
	</concept>
	<concept>
	<concept_id>10002978.10003022.10003028</concept_id>
	<concept_desc>Security and privacy~Domain-specific security and privacy architectures</concept_desc>
	<concept_significance>500</concept_significance>
	</concept>
	</ccs2012>
\end{CCSXML}

\ccsdesc[500]{Information systems~Email}
\ccsdesc[500]{Security and privacy~Security protocols}
\ccsdesc[500]{Security and privacy~Domain-specific security and privacy architectures}

\keywords{electronic mail, secure email, end-to-end encryption, privacy, usability}

\maketitle

{\itshape
\textsuperscript{\textasteriskcentered}%
This version extends our work published at Financial Cryptography 2021~\cite{clark2021sok}.}

% = = = = = Main Body = = = = = %

% !TEX root = ../main.tex

\section{Introduction}

Email has been called \textit{``probably the most valuable service on the Internet''}~\cite{bellovin2004look}. It has evolved over its 50-year history to become a pillar of seamless interoperability---if you know someone's email address, you can send email to them \cite{partridge2008technical} across a diverse range of desktop, mobile, and web client software. As an indication of its near-universal acceptance, an email address is often required to create online accounts and to make online purchases. As of 2020, there were an estimated 4 billion users of email sending over 306 billion email messages per day~\cite{radicati2020}. Despite its ubiquity, email was not created the security desirable for its ensuing wide deployment. 

Work to provide security for email, in various forms, has been ongoing for over three decades.
Early efforts focused on the confidentiality, authenticity, and integrity of email messages, with efforts to develop PEM~\cite{RFC1421} leading to work on S/MIME~\cite{RFC2633} and then, as a reaction, PGP~\cite{zimmermannpgp}.  
However, as measured in recent years, email is only sometimes transmitted over an encrypted connection, with limited protection from passive network eavesdropping and active network attacks~\cite{durumeric2015neither,foster2015security,mayer2016no,holz2016tls}.
Meanwhile, S/MIME has only seen limited uptake within enterprises and experts are abandoning PGP.\footnote{Including Phil Zimmermann~\cite{zimmermannpgp}, the creator of PGP; Moxie Marlinspike~\cite{marlinspikepgp}, who called PGP a ``\textit{glorious experiment that has run its course},'' and Filippo Valsorda~\cite{valsorda2016arstechnica}, who bemoans the  challenges of maintaining long-term PGP keys.}
Greater attention has focused on spam, malware, and phishing as they became problems for everyday users.
While spam filtering by many email providers has significantly improved, extensive email archives are typically stored in plaintext and vulnerable to hacking, and fraud through phishing and spear phishing remain problematic~\cite{rivner2011anatomy}.
It is within this context that we set out to systematically understand what went wrong with email security, how email security can theoretically be improved, and how tussling between stakeholders can lead to inaction.

\begin{table}[t]
	\def\arraystretch{1.5}
	
	\begin{tabularx}{\columnwidth}{lX}
		\toprule
		Stakeholder              & Description                                                                                                                                                                                                                             \\
		\midrule
		
		Email Service Providers  & Organizations that provide email services to industry and the public                                                                                                                                                            \\
		
		Enterprise Organizations & Large organizations in both government and industry                                                                                                                                                                                     \\
		
		Privacy Enthusiasts      & Users with strong privacy preferences who believe email should offer strong protection from corporate or government surveillance                                                                                                        \\
		
		Vulnerable Users         & Users who deal with strongly sensitive information that could induce personal safety risks, including journalists, dissidents, whistleblowers, informants, and undercover agents; we also include criminals as part of this stakeholder (due to aligned goals, despite ethical differences) \\
		
		Secure Mailbox Providers   & Organizations that provide secure email services to the public                                                                                                                                 \\
		
		Typical Users            & Users of standard, plaintext email services                                                                                                                                                                                             \\
		
		 Enforcement          & National security, intelligence, and law enforcement                                                                                                                                                                   \\
		
		\bottomrule
	\end{tabularx}
	\def\arraystretch{1}
	
\caption{Stakeholders with an interest in email and secure email.\label{stakeholders}}

\end{table}

\paragraph{Contributions and Methodology.}
To better understand the current state of affairs and identify where future research and development efforts should focus, we conduct a stakeholder-based analysis of secure email systems.
Our initial deliverable was a framework to evaluate secure email cryptographic primitives (\S\ref{sec:evaluations-crypto}), key management schemes (\S\ref{sec:evaluations-keymanagement}), and systems (\S\ref{sec:evaluations-systems}), allowing us to map out the landscape of solutions and compare how they satisfy a set of security, utility, deployability, and usability properties.
Ensuing discussion and review of this framework encouraged us to look specifically at how the actions and interests of a set of stakeholders (Table~\ref{stakeholders}) helps to explain the history of failures and successes in secure email, leading to the current patchwork of partial secure email solutions.
Using this new orientation for the paper, we systemize the academic literature on email, relevant IETF standards, industry solutions and software projects.
For each, we consider which stakeholder is behind the proposal, determine how it furthers the goals of the stakeholder, and infer how these goals compose with the goals of other stakeholders (\S\ref{sec:priorities}).
This allows us to identify incompatibilities, illustrate how different solutions have evolved to meet their needs, and show which stakeholders are under-served.

While we did not follow a standard or formal methodology for identifying research literature, our approach was as follows.
We (i) examined the proceedings of top ranked security, cryptography, and measurements venues; (ii) expanded the research set by contemplating other work that was cited in the papers we identified; and (iii) relied on our personal experience (which, for some, dates back to the early 1990s) and our acquired knowledge of the literature.
Similarly, the stakeholder groups were extracted from the literature through experience and discussion.
It is likely that a different set of authors would end up with a somewhat different set of papers and categorizations, but this seems to be true of nearly all SoKs at top security venues.

\paragraph{Rise of Secure Instant Messaging.}
The relatively low level of adoption of secure email is often contrasted with the wider success of secure messaging applications.
WhatsApp and Facebook Messenger have over a billion users, while iMessage, Signal, Telegram, Line, and Viber have millions.
The best of these provide forward secrecy and message deniability~\cite{borisov2004off,perrin2016doubleratchet} in addition to end-to-end encryption.
Unger et al.~\cite{unger2015sok} have an excellent systematization of secure messaging.
Yet, despite some calls to abandon secure email in favor  of Signal~\cite{valsorda2016arstechnica}, there are important reasons to not give up on email.
Email is an open system, in contrast to messaging's walled gardens, giving it fundamentally different uses, often involving longer messages, archival, search, and attachments.
There is no indication email is going away anytime soon.
As such, there is still an important need to increase the security and privacy of email-based communication.
% !TEX root = ../main.tex

\section{Preliminaries}
\label{sec:baseline}

\begin{figure}[t]
	\centering
	\includegraphics[width=\columnwidth]{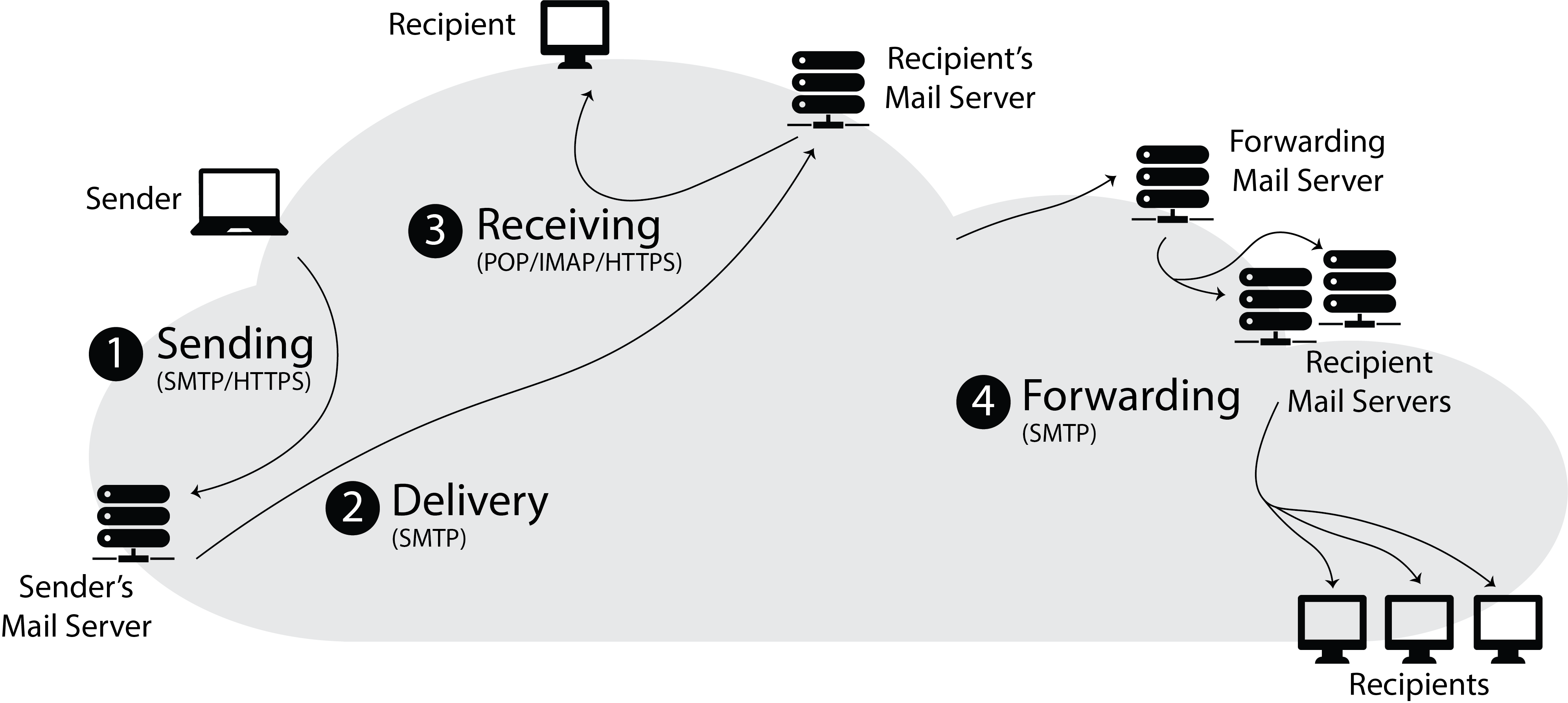}
	\caption{Overview of email operation and protocols. (1) Sending
		email generally uses SMTP or HTTPS between a client and its mail
		server. (2) Delivery of email between mail servers uses SMTP. (3)
		Receiving email generally uses POP, IMAP, or HTTPS. (4) Any mail server
		receiving email may forward it to other servers. This happens when a user asks to forward their email to a different account, or when a user sends to a mailing list.}
	\Description{Overview of email operation and protocols. (1) Sending
		email generally uses SMTP or HTTPS between a client and its mail
		server. (2) Delivery of email between mail servers uses SMTP. (3)
		Receiving email generally uses POP, IMAP, or HTTPS. (4) Any mail server
		receiving email may forward it to other servers. This happens when a user asks to forward their email to a different account, or when a user sends to a mailing list.}
	\label{fig:smtp}
\end{figure}

A series of protocols are used to send email, transfer it from the sender's email provider to the recipient's provider, and then retrieve it.
Figure~\ref{fig:smtp} shows the most basic steps involved, in steps marked (1) through (3). When a user initiates sending an email, their client may use SMTP~\cite{rfc5321} to submit the message to their organization's mail server (also called a mail transfer agent or MTA~\cite{rfc1506,rfc5598}).
The sender's MTA uses DNS to locate the mail MTA for the recipient's domain, then uses SMTP to transfer the message.
Finally, the recipient retrieves the message from their own organization's MTA, possibly using POP or IMAP.
If either the sender or receiver is using webmail, then step (1) or step (3) may use HTTPS instead.
Note also that the version of SMTP used to submit a message in step (1) is modified from the version of SMTP used to transfer messages~\cite{rfc6409}.

This sequence of events is complicated somewhat by additional features supported by email as shown in step (4). First, a receiving MTA can be configured to forward email for a recipient on to another MTA; \eg forwarding email from \texttt{bob@company.org} to \texttt{bob@gmail.com}.
This can repeat an arbitrary number of times.
Second, a destination email address may correspond to a mailing list server which forwards the email to all subscribers on the list  (a potentially large number).
This adds numerous other recipient MTAs to the process.

An email message itself consists of two parts: the envelope and the body. The envelope contains SMTP commands that direct MTAs regarding how the message should be delivered.
In particular, the envelope specifies the sender's email address (\texttt{MAIL FROM}) and the recipient's email address (\texttt{RCPT TO}).
The message body has a separate format, including the familiar \textit{From}, \textit{To}, \textit{CC}, and \textit{Subject} header fields.
Email clients generally display the sender's email address shown in the \textit{From} header in the body, rather than the one in the SMTP envelope. 

\paragraph{Why Email is Insecure.} Every aspect of email was initially designed, specified, and developed without foreseeing the need for security protection that would later be recognized given how universal email has become. Security issues persist today despite decades of work to fix them. The original designs of protocols used to send, receive, and deliver email among clients and servers contained no protections for integrity or confidentiality. All messages were transmitted in the clear and could be intercepted and modified by anyone able to act as a man-in-the-middle. The original specifications contain nothing that validates the \texttt{MAIL FROM} command or prevents forgery of the \textit{From} header. The ease of forging emails did nothing to inhibit the emergence of unsolicited email. Email never easily faciliated network-level anonymity, message deniability, or untraceability.

% !TEX root = ../main.tex

\section{Stakeholders}
\label{sec:stakeholders}
 
The premise of our systematization of knowledge is that understanding the tussles among stakeholders are central to understanding why secure email lacks a universal solution. We identified potential stakeholders through an extensive period of analysis that included reviewing the research literature; reading online posts, discussion threads, and news articles regarding secure email; and by looking at press releases and features provided by secure email tools. We then carefully distilled the set to \textit{key} stakeholders who: (1) reflect unique preferences, and (2) are important to the history of research and development in this area (see Table~\ref{stakeholders}).

An example of a stakeholder that is not a key stakeholder within our framework would be a company that produces client email software, as these companies tend to reflect the preferences of their customers---customers that are already key stakeholders like \textit{enterprise organizations}, \textit{typical users}, and \textit{privacy enthusiasts}. Another example is government which is multifaceted. Many government departments operate like \textit{enterprise organizations}, while others are captured by \textit{enforcement}. But even within national security, law enforcement and intelligence agents and assets themselves have the preferences of \textit{privacy enthusiasts} or \textit{vulnerable users}.  In this section, we align various efforts toward secure email with the appropriate stakeholders and in Section~\ref{sec:priorities} discuss the trade-offs. 

% = = = = = = = = = = = = = = = = = = = = = = = %
% = = = = = = = = = = = = = = = = = = = = = = = %
% = = = = = = = = = = = = = = = = = = = = = = = %

\subsection{Email Service Providers}
\label{sec:providers}

An email service (or mailbox) provider~\cite{rfc5598} is focused on retaining its customers for business and personal use. Providers have adopted several technologies to improve the security of email, including link encryption, domain authentication, and sender
authentication. Providers often require access to plaintext so they can scan incoming emails for spam and malware. We review current and planned efforts, the protection they offer, and assessments of their effectiveness.

\subsubsection{Link Encryption.} 
\label{sec:link-encryption}

Providers have adopted methods for encrypting email while it is in transit between MTAs or between an MTA and a client. Such `link' encryption is designed to prevent eavesdropping and tampering by third parties that may own untrusted routers along the path that email is being delivered~\cite{rfc3207}, however messages are not protected from inspection or modification at each MTA. While more privacy invasive than end-to-end encryption (encryption between the email sender and recipient), link encryption enables providers to scan for malicious email attachments, classify potential spam or phishing attacks, modify email tracking links, and provide other services.

Mail transferred with SMTP between MTAs is by default plaintext, and an MTA can use the STARTTLS command~\cite{rfc3207} to negotiate an encrypted channel. However, an active adversary between the MTAs can corrupt or strip STARTTLS, downgrading the connection to plaintext~\cite{durumeric2015neither}. A recent initiative (currently called MTA-STS~\cite{ietf-margolis-smtp-sts-00}) provides a way for an MTA to advertise a strict transport security (STS) policy stating that they always require STARTTLS. The policy is trusted on first use (TOFU) or authenticated using the certificate authority (CA) system. Should DNSSEC become widely deployed, policies can be directly advertised by the MTA in its DNS record~\cite{barnes2011dane,rfc6698}. Even with link encryption, SMTP reveals significant metadata about email messages---some proposed mitigations have been drafted~\cite{dime,sparrow2016leap}. 

Recall that email client software most often uses IMAP (or the older POP3) to retrieve mail and SMTP to send messages. STARTTLS is supported across each of these protocols~\cite{rfc2595} and is often required by the mail server. Users of webmail typically access their mail client using HTTPS. Under the link encryption paradigm, end users can ensure encryption to their mail server but have no control over (or even visibility of) the use of encryption for the transport of their emails. 

\subsubsection{Authentication.} Consider the case when Alice receives an email from \texttt{bob@gmail.com}. \emph{Domain authentication} indicates that the email was sent by a server authorized to send email from \texttt{gmail.com}, while \emph{sender authentication} validates the user account \texttt{bob@gmail.com} originated the mail. The final level of authentication is \textit{user authentication}, which occurs when Alice ensures that a human, such as Bob Smith owns the \texttt{bob@gmail.com} account. While user authentication is ideal, it taps into a public key infrastructure that email providers have avoided, settling instead for \textit{domain authentication}, which has a long history rooted in identifying spam and filtering
malware~\cite{rfc7208,rfc4686,rfc7489,ietf-dmarc-arc-protocol-08}.

\subsubsection{Domain Authentication.}
\label{sec:domain-auth}

The primary protocol for domain authentication is DomainKeys Identified Mail (DKIM)~\cite{rfc5585,rfc6376}. The server originating email for a particular domain will generate a digital signature key pair, advertise the public key in the DNS record for the same domain, and sign all outbound email, with the appropriate validation data added to a header field in the email. A well-positioned adversary can modify a recipient's retrieval of the public key from DNS---DNSSEC can mitigate this threat~\cite{rfc4033}. DKIM signatures are fragile to any modification to the message body or header fields.

Using the same principle of advertising through DNS records, Sender Policy Framework (SPF)~\cite{rfc7208} allows a domain
to specify which IP addresses are allowed to originate email for their domain, while Domain Message Authentication, Reporting, and Conformance (DMARC)~\cite{rfc7489} enables specification of which services (DKIM, SPF) they support, along with a policy
indicating what action should be taken if authentication fails. DMARC has many additional features around reporting misconfigurations and abuse, but importantly it also requires identifier alignment. For SPF, this means that the domain in the envelope \texttt{MAIL FROM} address (which is authenticated with SPF) must match the domain in the \textit{From} header field. For DKIM, this means that the domain used for signing must match the domain in the \textit{From} header field. This links the authentication or signature verification done by SPF and DKIM to the \textit{From} address seen by the user.

Security extensions like SPF and DKIM were developed at different times for different purposes. DMARC is intended to cover gaps between SPF and DKIM. Such a patchwork approach to security is often susceptible to vulnerabilities, particularly when the protocols are implemented across different client and server software components that need to interoperate. A recent study on the composition of SPF, DKIM, and DMARC identifies 18 attack vectors and finds that all tested mail providers and email clients were vulnerable to at least one~\cite{CPJ20}. 

\subsubsection{Sender Authentication.}
\label{sec:senderauth}

There is no wide support for sender authentication. Most mailbox providers do authenticate their users~\cite{imcr016}. For example, if the sender is using webmail, then she may authenticate by logging into her webmail account. If the sender is using a desktop client, the mail domain can authenticate her with SMTP Authentication, which provides several methods that enable the sender to authenticate with the MTA
by a username and password~\cite{rfc4954,rfc5034,rfc4959}. However, the measures a domain uses to authenticate a sender are not
communicated to the recipient of an email message, nor can they be verified by the recipient.

\subsubsection{Reducing the Fragility of Authentication.} 

Authenticated Received Chain (ARC)~\cite{ietf-dmarc-arc-protocol-08,ietf-dmarc-arc-usage-02} extends email authentication to handle cases when messages are potentially modified when being forwarded, such as by a mailing list. With ARC, authentication checks are accumulated by forwarders in a message header field~\cite{rfc8601} as well as a signature on the email as received (these header fields are sealed with an additional signature by each forwarder, creating a chain). The protocol is intended for broad use by all email handlers along a transmission path, not just perimeter MTAs, and it is designed to allow handlers to safely extend the chain even if when they are certain they have not modified the message. When all email handlers are trusted by the recipient, ARC enables any modifications to the message to be attributed, and for DKIM, SPF, and DMRAC results to be validated on the pre-modified message. However, a malicious handler is not prevented from altering messages or removing ARC headers.

 \subsubsection{Mitigating Email Misuse.}
 \label{ref:spam}
 
Mailbox providers have invested significant effort in spam, phishing, and malware filtering.  In the early 2010s, a successful malicious email campaign might see a spammer employ a botnet of 3,000 compromised machines to send 25 billion emails to 10 million addresses~\cite{iedemska2014tricks}. Each aspect of the pipeline---from the compromised machines to the email list to the software tools---might be sold by specialists~\cite{levchenko2011click}, and the campaign itself is typically run by third-party advertisers earning pay-per-click revenue for directing traffic to a third-party site (\eg  storefronts for unregulated pharmaceuticals constitute over a third of spam)~\cite{MPG+21}. 
 
Spam filtering has evolved from IP address blacklists to highly sophisticated classifiers that examine content,  meta-information including origin, user reports, and protocol details such as SMTP header fingerprints \cite{stringhini2012b}. Malware filtering is often performed by comparing email attachments to signatures of known malware.  Spammers use a variety of evasion techniques, including sending from the IP addresses of  malware-compromised computers~\cite{franklin2007inquiry}, spoofing sender addresses, and encoding text as images. An  esoteric proposal for spam prevention is requiring the sender to compute a costly function to send an email~\cite{dwork1992pricing,back2002hashcash}---an approach that never caught on~\cite{laurie2004proof}.

\subsubsection{Measurement Studies of Adoption and Effectiveness.}
\label{sec:measurement}

In 2015--2018, several papers were published~\cite{durumeric2015neither,foster2015security,mayer2016no,holz2016tls,HW18} that measured the level of adoption and effectiveness of the encryption and domain authentication used by email providers. The general picture they paint is that top email providers encrypt messages with STARTTLS and use SPF and DKIM for authentication, but there is a long tail of organizations that are lagging in deploying these mechanisms. However, even when protection methods within email are deployed, they are often compromised by insecure practices, such as acceptance of: self-signed certificates\footnote{With the advent of free domain certificates with Let's Encrypt, it is possible that more providers are using verifiable certificates.} (when CA-signed certificates were expected), expired certificates, or broken chains, all of which cause the validation of the certificate to fail. Email traffic often uses weak cipher suites, weak cryptographic primitives and parameters, weak keys, or password authentication over unencrypted connections. Of the techniques that rely on DNS, basic attacks such as DNS hijacking, dangling DNS pointers~\cite{LHW16}, and modifying non-DNSSEC lookups can enable circumvention. Stripping attacks can compromise STARTTLS, with Durumeric et al.~\cite{durumeric2015neither} illustrating how these attacks caused 20\% of inbound Gmail messages to be sent in cleartext for seven countries. Use of SPF is common, but enforcement is limited, and DNS records often are not protected with DNSSEC.  There is little use of DKIM, and few servers reject invalid DKIM signatures~\cite{foster2015security}.

As Mayer et al. ~\cite{mayer2016no} conclude, \textit{``the global email system provides
		some protection against passive eavesdropping, limited protection
		against unprivileged peer message forgery, and no protection against
		active network-based attacks.''}

% = = = = = = = = = = = = = = = = = = = = = = = %
% = = = = = = = = = = = = = = = = = = = = = = = %
% = = = = = = = = = = = = = = = = = = = = = = = %

\subsection{Enterprise Organizations}
\label{sec:enterprise}

Enterprises have overlapping interests with email service providers (like reducing email misuse) but often prefer stronger (end-to-end) encryption and authentication, at least within their internal boundaries. Enterprises played a role in developing standards that could meet their needs, starting with PEM~\cite{kent1993internet,RFC1421,RFC1422,RFC1423,RFC1424} and leading to S/MIME~\cite{RFC2633,RFC5751,RFC5280}. Another issue that is highly relevant to enterprises is mitigating carefully targeted social engineering attacks against its employees, often conducted through email. 

\subsubsection{End-to-end encryption and authentication.}
\label{sec:smime}

The primary goals of PEM~\cite{kent1993internet,RFC1421,RFC1422,RFC1423,RFC1424} were end-to-end email security with confidentiality, data origin authentication, connectionless integrity (order not preserved), non-repudiation with proof of origin, and transparency to providers and to SMTP. PEM was distinguished by interoperability with non-PEM MTAs, and a hierarchical X.509 public key infrastructure (PKI) with revocation that largely precludes rogue certificate issues haunting later PKI systems. A contributing factor cited \cite{orman2015encrypted} in PEM's demise was its slow progress in evolving for Multipurpose Internet Mail Extensions (MIME)~\cite{RFC2045}, the standard for including attachments, multi-part bodies, and non-ASCII character sets. Industry support moved to S/MIME, while privacy advocates favored PGP (see Section~\ref{sec:privacy-focused}) because it was free from the restrictions imposed by PEM's centralized and hierarchical organization.

S/MIME~\cite{RFC5751} is a standards suite for securing MIME data with both encryption and digital signatures. It was originally developed during the early 1990s by RSA Data Security, then later adopted by the IETF, resulting in standards in 1999~\cite{RFC2633,RFC5751,RFC5280}. S/MIME's Cryptographic Message Syntax (CMS)~\cite{rfc5652} has origins in PEM and PKCS. S/MIME has wide support on major platforms and products~\cite[p.60--62]{orman2015encrypted}. S/MIME clients use \textit{certificate directories} to look up X.509v3 certificates.\footnote{Of note, S/MIME uses a supporting suite of certificate management protocols, including RFC 5280~\cite{RFC5280}, which defines an IETF subset of X.509v3 certificates.} S/MIME does not mandate a hierarchy with a single root certificate authority (CA) and any organization can act as an independent, trusted root for its certificates---the most common usage today. Interoperability between organizations is limited or non-existent.
%remains in advancement~\cite{oasis2016key}.

Several works have examined usability deficiencies with S/MIME implementations, noting difficulties knowing which CAs to trust~\cite{kapadia2007case}, difficulties with certificate management~\cite{fry2012not}, and inconsistency in handling certificates~\cite[p.60--67]{orman2015encrypted}. Automatically creating and distributing signing and encryption keys at account creation is considered good practice~\cite{garfinkelCHI2005}.

%\subsubsection{Automating key use.}
%\label{sec:hosted}

\subsubsection{Private Key Escrow.}
\label{sec:hosted}

Enterprises often use \textit{private key escrow} in conjunction with S/MIME, which enables the organization to decrypt emails and scan for spam, malware, fraud, and insider trading, as well as archiving messages for regulatory reasons and enabling recovery if a client loses its private key. The suitability of S/MIME's centralized certificate management for enterprises and government has led to large, but siloed, deployments~\cite{chandramouli2016trustworthy}.
Some providers simplify S/MIME deployment using \emph{hosted S/MIME}~\cite{hostedsmime}, where an enterprise uploads user private keys to an email provider, and the provider automatically uses S/MIME for some emails (\eg to other users of the same provider). Encryption in this case is only \textit{provider-to-provider} rather than end-to-end.

% JC: should add Boneh-Franklin as Shamir et al is just a sketch of how it might be done without a concrete implementation

\label{sec:ibe}
As an alternative to S/MIME, some enterprise email solutions rely on identity-based encryption (IBE)~\cite{shamir1984identity}.
IBE uses a trusted server to store a master private key and generate individual private keys for users.  The trusted server also advertises a master public key, which clients can use to derive a public key for any email address.  Users can validate their ownership of an email address with the IBE server to retrieve their generated private key. IBE simplifies key management for clients but leaves the IBE server with persistent access to each user's private key, and also substantially complicates revocation~\cite{boldyreva2008identity}.\footnote{Revocation of a compromised private key can be supported by having versions of the key. The result of obtaining an incorrect key version is comparable to obtaining a compromised key. The trust model of IBE is tantamount to a trusted public key server.} Ruoti et al.~\cite{ruoti2013confused,ruoti2016private} integrated IBE into a webmail system, demonstrating how automating interactions with key management results in successful task completion and positive user feedback.

\subsubsection{Transparent email encryption.}
A distinct approach to making interactions with PKI transparent to users is to layer encryption and signing below client software. Levien et al.~\cite{levien1996transparent} places this functionality between the email client software and the MTA, while Wolthusen~\cite{wolthusen2003distributed} uses the operating system to intercept all network traffic and then automatically apply email encryption. Currently, several companies (\eg Symantec) offer automated encryption of emails by intercepting them as they traverse a corporate network.

\subsubsection{Spear Phishing.}

Social engineering may be crafted as a generic attack but is often a targeted attack against specific enterprise employees. The openness of email enables direct contact with targets and an opportunity to mislead the target through the content of the email, a spoofed or
 look-alike send address, and/or a malicious file attachment~\cite{mitnick2011art,Had10}. As an illustration, the company RSA was breached through a sophisticated attack that started with a targeted email impersonating an employee and a corrupted spreadsheet attachment~\cite{rivner2011anatomy}. Employee training~\cite{caputo2014going} and email filtering are important countermeasures, however spam filters are typically trained to detect \emph{bulk} email delivery and classifying bespoke spear phishing emails remains a challenge~\cite{laszka2015optimal}. 
 
% = = = = = = = = = = = = = = = = = = = = = = = %
% = = = = = = = = = = = = = = = = = = = = = = = %
% = = = = = = = = = = = = = = = = = = = = = = = %

\subsection{Privacy Enthusiasts}
\label{sec:privacy-focused}

Privacy enthusiasts prefer end-to-end encrypted email to avoid government surveillance or commercial use of their data generally.
They differ from vulnerable users (see section~\ref{sec:vulnerable}) in that there is not an immediate personal safety risk driving their usage of secure email. Privacy enthusiasts have historically favored PGP, which was developed as ``public key cryptography for the masses" and ``guerrilla cryptography" to counter authorities~\cite{zimmermann1995official}. The difficulty with PGP has always been finding a suitable alternative to the centralized trust model of S/MIME. 

\subsubsection{End-to-end encryption and authentication.}
\label{sec:pgp}

PGP's history is a fascinating 25-year tale of controversy, architectural zig-zags, name ambiguity, and patent disputes,
with changes in algorithms, formats and functionality;
commercial vs.\ non-commercial products; corporate brand ownership; and circumvention
of U.S. crypto export controls.\footnote{PGP was distributed
	as freeware on the Internet in 1991, leading to an investigation of Zimmermann
	by the United States Customs Office for allegedly violating U.S. export laws.
	He published the PGP source code in book form in 1995~\cite{zimmermann1995pgp},
	and the case was subsequently dropped in 1996~\cite{lauzon1998philip}.}
The current standard for the PGP message format is OpenPGP~\cite{RFC4880,rfc3156},
a patent-unencumbered variation. Despite evolving formats or encryption algorithms,
PGP enthusiasts until recently have largely remained faithful to
PGP's distinguishing concepts:

\begin{itemize}

\item \textbf{PGP key packets and lightweight certificates:} 
PGP key packets hold bare keys (public or private). Public keys are kept in \textit{lightweight certificates} (compare~\cite{zimmermann1995official}), which are not signed certificates in the X.509 sense, but instead contain keys and a User ID (username and email address). To help client software determine which keys to trust, PGP also includes \textit{transferable public keys}~\cite{RFC4880}, which include one or more \textit{User ID packets} each followed by zero or more \textit{signature packets}. The latter attest the signing party's belief that the public key belongs to the user denoted by the User ID. Users typically store private keys on their local device, often encrypted with a password, though hardware tokens are also available.
	      
\item \textbf{PGP's web of trust:} \label{sec:web-of-trust}
The web of trust (WoT) is a model in which users personally decide whether to trust public keys of other users, which may be acquired through personal exchanges or from public servers, and which may be endorsed by other users they explicitly designate to be \textit{trusted introducers}~\cite{zimmermann2011pgp}.
	      
\item \textbf{PGP key packet servers:} \label{sec:pgp-server}
Users publish their public key to either closed or publicly accessible key packet servers, which contain a mapping of email address to the public key. Clients query to locate the public key associated with an email address.
	      
\end{itemize}

\subsubsection{Problems with PGP.}

PGP's design around the web of trust has allowed quick deployment in small groups without bureaucracy or costs of formal Certification Authorities~\cite{mcgregor2017panama}, but leads to other significant obstacles:  

\begin{itemize}

\item \textbf{Scalability beyond small groups:}
Zimmerman notes~\cite[p.23]{zimmermann1995official} that \textit{``PGP was originally designed to handle small personal keyrings"}. Scaling PGP requires acquiring large numbers of keys, along with a manual trust decision for each key, plus manual management of key storage and the key lifecycle.
	      
\item \textbf{Design failure to address revocation:} Zimmermann writes~\cite[p.31]{zimmermann1995official}, \textit{``If your secret key is ever compromised...you just have to spread the word and hope everyone hears about it".}  PGP does have methods to revoke keys, but distribution of these to others is ad hoc.
	      
\item \textbf{Usability by non-technical users:} Zimmerman~\cite[p.31]{zimmermann1995official} says \textit{``PGP is for people who prefer to pack their own parachutes"}. There is no system help or recovery if users fail to back up their private key or forget their passphrase. Furthermore, users must understand the nuances of generating and storing keys, trusting public keys, endorsing a public key for other users, and designating others as trusted introducers. The poor usability of PGP has received significant attention~\cite{whitten1999why,ruoti2019usability}.
	      
\item \textbf{Trust model mismatch:} Zimmerman notes~\cite[p.25]{zimmermann1995official} that \textit{``PGP tends to emphasize [an] organic decentralized non-institutional approach"} reflecting personal social interaction rather than organizational relationships. The PGP web of trust was designed to model social interaction, rather than decision-making processes in governments and large enterprises. It is thus not a one-size-fits-all trust model.
	      
\end{itemize}

\subsubsection{Trust-on-first-use (TOFU).}
\label{sec:tofu}

An alternative to PGP's web of trust is to exchange keys in-band and have clients trust them on first use. This has been the subject of several research projects~\cite{roth2005security,garfinkel2005johnny,masone2009abuse}. Since 2016, the developer community has been integrating TOFU into PGP implementations in the MailPile, PEP~\cite{birk-pep-00}, LEAP~\cite{sparrow2016leap}, and Autocrypt~\cite{autocryptspec} projects. A common critique of TOFU is that users cannot distinguish valid key changes from an attack.  Recent work by developers in the PEP and LEAP projects is aiming to address this problem with additional methods to authenticate public encryption keys, such as using a trusted public key server, auditing public key servers, and the fraught procedure of asking the user to compare key fingerprints~\cite{hsiao2009study,dechandempirical}.

\subsubsection{Public key servers and logs.}
\label{sec:trusted}
\label{sec:audited}

Another web of trust alternative---applicable to (and aligned with) S/MIME's trust model---is introducing a trusted public key server. Recent work~\cite{atwater2015leading,ruoti2018comparative} showed that automated servers have high usability when integrated into a PGP-based email system.  Bai et al.~\cite{bai2016inconvenient} found users prefer key servers to manual key exchange, even after being taught about the security limitations of a key server.

A compromise between TOFU and a fully trusted server is to allow key assertions from users but ensuring they are published publicly in untrusted logs, allowing monitors to examine a history of all certificates or key packets that a key server has made available for any entity~\cite{ryan2014enhanced,melara2015coniks,basin2015arpki}. This enables detection of rogue keys and server equivocation.

\subsubsection{Social Authentication.}

Another way to disseminate public keys is to associate them with public social media accounts. The Keybase project\footnote{\url{https://keybase.io}}
helps users to post a signed, cryptographic proof to their account, simultaneously demonstrating ownership of a public key
and ownership of the account. By aggregating proofs across multiple social media accounts for the same person, a client can establish
evidence that ties a public key to an online persona, under the assumption that it is unlikely that a person's social media
accounts are all compromised simultaneously. The Confidante email system leverages Keybase for
distribution of encryption public keys, with a study finding it was usable for lawyers and journalists~\cite{lerner2017confidante}.

\subsubsection{Short-lived keys and forward secrecy.}

% JC: If we need space, this would be a candidate for removal (theoretical more than practical). Also seems to blend IM with email in a confusing way (email has no interaction so things like OTR, etc are not possible. Puncturable encryption was a fad

Schneier and Hall~\cite{schneier1997improved} explored the use of short-term private keys to minimize the damage resulting
from the compromise of a private key. Brown and Laurie~\cite{brown2000security} discuss timeliness in destroying a short-lived
key and how short-lived keys complicate usability by requiring more frequent key dissemination. %Off-the-Record Communication~\cite{borisov2004off} expanded on this vision with a protocol that provides forward secrecy on instant messaging platforms and can be applied to encrypt all but the initial email message sent between two parties. This work led to the double ratchet algorithm used in Signal~\cite{perrin2016doubleratchet}. Puncturable encryption~\cite{green2015forward} also provides forward secrecy that can be used for email and allows a recipient to revoke decryption capability for specific messages or time periods.

% = = = = = = = = = = = = = = = = = = = = = = = %
% = = = = = = = = = = = = = = = = = = = = = = = %
% = = = = = = = = = = = = = = = = = = = = = = = %

\subsection{Vulnerable Users}
\label{sec:vulnerable}

Vulnerable users deal with strongly sensitive information that could induce personal safety risks. Using email from a malware-infected device is a primary concern~\cite{chatterjee2018ipv,havron2019ipv}, as well as risks due to the design and common practices of email.

\subsubsection{Pseudonymity.}
\label{sec:layered}

One concern for vulnerable users is the inability to forgo leaking personally identifiable meta-information: \ie unlink the contents of the email from their true email address, their IP address, and/or the identity of their mail server. Technically inclined vulnerable users generally opt for pseudonymity~\cite{Gol00} where more than one email sent from the same pseudonymous account can be established as having the same origin, but no further information is known. 

Historically, PEM accommodated anonymous users with \textit{persona certificates}, which could provide assurances of continuity of a pseudonymous user ID but does not prevent network level traceability. Today, \textit{layered encryption} is used in which messages are routed through multiple non-colluding servers, with each server unwrapping a layer of encryption until the message is delivered to its destination, with the same happening for replies in reverse. This idea was championed by the cypherpunk movement~\cite{Nar13a} and adapted to the email protocol with remailers like mixminion and others~\cite{GWB97,Gol03,Gol07,DDM03}. Pseudonymity is realized as indistinguishability from a set of plausible candidates---the set of other users at the time of use~\cite{DM06}---which may be small, depending on the system and circumstances.\footnote{To illustrate, a student emailed a bomb threat to Harvard's administration via webmail accessed over Tor~\cite{dms04}. The suspect was found to be the only individual accessing Tor on Harvard's network at the time the email was sent---while strictly circumstantial, the suspect confessed~\cite{goodin}.}

A simpler approach is to register a webmail account under a pseudonymous email address, optionally using Tor~\cite{dms04} to access the mailbox. Satoshi Nakamoto, the inventor of Bitcoin~\cite{Nak08}, corresponded over webmail for many months while remaining anonymous.

\subsubsection{Traceability, deniability, and ephemerality.} 
\label{sec:ephemerality}

Email senders for some time have abused the browser-like features of modern email clients to determine when recipients view an email, when a links are clicked, and (via third-party trackers) what other collected information is known about the recipient~\cite{nar18}. Email service provider interventions can interfere with domain authentication (DKIM).

Deniability considers a case where the recipient wants to authenticate the sender, but the sender does not want the evidence to be convincing to anyone else. Cryptographers have suggested new signature types~\cite{Cha95,JSI96,RST01} to provide deniability, but these typically require trusted third parties and/or a robust PKI and have near-zero deployment.

Once sent, a sender loses control over an email and the extent to which its contents will be archived. In order to automate a shorter retention period, emails might contain a link to the message body which is deposited with and automatically deleted by a trusted service provider or a distributed network~\cite{GKLL09,WHHF+10}.

% = = = = = = = = = = = = = = = = = = = = = = = %
% = = = = = = = = = = = = = = = = = = = = = = = %
% = = = = = = = = = = = = = = = = = = = = = = = %

\subsection{Secure Mailbox Providers}
\label{sec:webmail}

A secure mailbox provider offers end-to-end encryption and authentication between users of their service. Providers like ProtonMail~\cite{protonmail}, Hushmail~\cite{hushmail}, and Tutanota~\cite{tutanota} have millions of users combined. Users' private keys are password-protected client-side and then stored with the provider, preventing provider access (assuming the password is strong~\cite{florencio2014administrator}) while allowing cross-device access. However, providers are trusted in other regards: inter-user encryption and authentication is generally blackbox and not independently verifiable,\footnote{Fingerprint comparison is common with secure messaging applications, but the feature is often ignored by users~\cite{schroder2016signal}.} and the model relies on client-side scripting where malicious (first or third-party) scripts would compromise security. Additional methods are needed to provide code signing and privilege separation for JavaScript in the browser~\cite{meyerovich2010conscript,webjail}. Generally, email sent to outside users are encrypted client-side with a one-time use passphrase, deposited in message repository with an access link sent as the original email (the passphrase is communicated between the sender and recipient out-of-band). 

A second approach is to use a browser extension to overlay signed and encrypted email on an existing mailbox provider. Initiatives here include automating PGP key management tasks (Mailvelope and FlowCrypt), providing automated S/MIME-based encryption and signing (Fossa Guard), encryption with a symmetric key held by the service (Virtru), or encryption using a password shared out of band
(SecureGmail). Google developed E2EMail to integrate OpenPGP with Gmail in Chrome but the project has been inactive for several years.

% = = = = = = = = = = = = = = = = = = = = = = = %
% = = = = = = = = = = = = = = = = = = = = = = = %
% = = = = = = = = = = = = = = = = = = = = = = = %

\subsection{Typical Users}
\label{sec:users}

It is well-known that most users do not use end-to-end encrypted email. For example, a recent evaluation of email usage at a large university found that over 27 years, only 0.06\% of users had encrypted email~\cite{stransky22}. Some work has examined the question of why most people do not use encrypted email. Renaud et al.~\cite{renaud2014doesn} found support for four reasons for non-adoption---lack of concern, misconceptions about threats, not perceiving a significant threat, and not knowing how to protect themselves. An earlier survey of 400+ respondents by Garfinkel et al.~\cite{garfinkelCHI2005} found that half indicated they didn't use encrypted email because they didn't know how, while the rest indicated they didn't think it was necessary, didn't care, or thought the effort would be wasted. Other work reports that users are unsure about when they would need secure email~\cite{ruoti2016we} and are skeptical that any system can secure their information~\cite{ruoti2017weighing,dechand2019encryption}. It
is not clear that users want to use digital signatures or encryption for daily, non-sensitive
messages~\cite{farrell2009ieee,gaw2006secrecy}.
Overall, work in this area demonstrates that usability is not the only obstacle to adoption, and that users don't perceive significant risk with email, lack knowledge about effective ways to mitigate risk, and don't have self-confidence
about their ability to effectively use secure systems.

The usable security and privacy community is increasingly utilizing new approaches
to address broader questions of adoption of security and privacy practices. Users are often rational when making decisions about whether to follow security advice; Herley~\cite{herley2009so} makes the case that users sometimes understand risks better than security experts, that worst-case harm is not the same as actual harm, and that user effort is not free. Sasse~\cite{sasse2015scaring} has likewise warned against scaring or bullying people into doing the ``right'' thing.
As a result, effort is being made to understand users' mental models~\cite{wash2010folk,fagan2016they,kang2015my,wu2018tree} when they
interact with secure software and using risk communication techniques to better
understand adoption or non-adoption of secure software~\cite{stewart2012death,wu2019something}, among other methods.

% = = = = = = = = = = = = = = = = = = = = = = = %
% = = = = = = = = = = = = = = = = = = = = = = = %
% = = = = = = = = = = = = = = = = = = = = = = = %

\subsection{Enforcement}
\label{sec:law-enforcement}

We broaden the term  enforcement to encompass police and law enforcement agencies, as well as national security and intelligence services. Law enforcement prioritizes access to plaintext communications, either through broad surveillance or exceptional access such as with a warrant. This need for access to plaintext communications has led to calls for so-called encryption back doors, leading to regular debates on whether this is desirable or feasible. This debate originally surfaced in the U.S. in the 1990s concerning email and has been rekindled regularly, now with greater emphasis on instant messaging which has seen better success than email at deploying end-to-end encryption to regular users. Proponents cite fears that widespread use of end-to-end encryption will enable criminals and terrorists to ``go dark'' and evade law enforcement. In response, privacy advocates decry growing mass surveillance, point to a history of abuses of wiretapping~\cite{Diffie:2007:PLP:1296070}, and suggest that market forces will ensure there is plenty of unencrypted data for use by law enforcement regardless~\cite{gasser2016don}.

A 2015 paper from Abelson et al.~\cite{abelson2015keys} highlights risks of regulatory requirements in this area, reiterating many issues discussed in their earlier 1997 report~\cite{abelson1997risks}. Identified risks include reversing progress made in deploying forward secrecy, leading to weaker privacy guarantees when keys are compromised; substantial increases to system complexity, making systems more likely to contain exploitable flaws; and the concentration of value for targeted attacks. Their report also highlights jurisdictional issues that create significant complexity in a global Internet. More broadly, whenever service providers have access to keys that can decrypt customer email, this allows plaintext to be revealed due to incompetent or untrustworthy service providers, by disillusioned employees, by government subpoena, or by regulatory coercion.

% !TEX root = ../main.tex

\section{Stakeholder Priorities}
\label{sec:priorities}

In the previous section, we aligned efforts toward secure email with their appropriate stakeholders.
In this section, we establish seventeen properties of secure email that are a priority to at least one stakeholder.
We first define these properties, then finish the section by ranking the importance of these properties to each stakeholder
These priorities are a result of extensive discussion among the authors using our literature review and current practices as evidence for our ratings.

\subsection{Property Definitions}
\label{sec:properties}

The seventeen properties can be split into four categories---security, utility, deployability, and usability.
% For ease of reference, each property is assigned a two-character designation (\eg S1, S2, D1) and a shorthand reference.
% While this approach is derived from the work of Bonneau et al.~\cite{BHOS12}, our properties and definitions are distinct from that work.
Below we define these seventeen properties and indicate what it means to have full or partial support for the property in a secure email system.

\subsubsection{Security Properties (\ref{col:confidentiality}--\ref{col:phishing})}\hfill

\col{S}{confidentiality}{Protection from eavesdropping}
\subparagraph{\ref{col:confidentiality} \ref{col-text:confidentiality}}
Full support indicates that the body of an email is kept confidential from all parties other than the sender and recipients.
Partial support indicates that the sender's and recipient's mail servers are able to read the body of the email, but not other relaying servers or network middleboxes.
This property does not reference whether envelope metadata and message headers, including subject lines, are protected.

\col{S}{integrity}{Protection from tampering and injection}
\subparagraph{\ref{col:integrity} \ref{col-text:integrity}}
Full support indicates that email is authored by the claimed sender and has not been modified by other parties.
Partial support indicates undetected modification or injection is possible by the sender's or recipient's mail servers.

\col{S}{private-key}{Private keys only accessible to user}
\subparagraph{\ref{col:private-key} \ref{col-text:private-key}}
Full support indicates that no third-party has access to a user's private signing and/or decryption private keys.
Partial support indicates password-encrypted private keys are stored by a third-party.
Note, many user-chosen passwords may be unlikely to resist an offline guessing attack.

\col{S}{exceptional-access}{Prevents exceptional access}
\subparagraph{\ref{col:exceptional-access} \ref{col-text:exceptional-access}}
Full support indicates exceptional access by corporations or law enforcement to
the plaintext contents of messages in exceptional circumstances is not possible.
Partial support indicates exceptional access is possible, but requires the cooperation or mutual-coercion of multiple parties (\eg email provider and key server).
A rating of no support indicates that exceptional access is possible and can be conducted by a single party (\ie does not require the cooperation/mutual-coercion of multiple parties).
For this property, we do not distinguish how exceptional access is enabled and whether or not a visible audit record is left as evidence, just that exceptional access is granted.
%Regular access to plaintext (\eg to scan for malware) implies exceptional access.

\col{S}{revocation}{Responsive public key revocation}
\subparagraph{\ref{col:revocation} \ref{col-text:revocation}}
Full support indicates an architecture allows for immediate and automatic checking of up-to-date revocation information.

\col{S}{public-key-audit}{Provides a public key audit trail}
\subparagraph{\ref{col:public-key-audit} \ref{col-text:public-key-audit}}
Full support indicates an audit log of public keys is available that provides non-equivocation~\cite{melara2015coniks}, so that impersonation attacks
can be detected.

\col{S}{privacy}{Supports sender pseudonymity}
\subparagraph{\ref{col:privacy} \ref{col-text:privacy}}
Full support indicates that it is generally infeasible for the recipient to learn the sender's IP address, the sender's email address, and the sender's mail server.
Note, because the contents of a message could potentially de-anonymize the sender, end-to-end encryption of the email body (\ref{col:confidentiality}) and the subject should also be used when desiring pseudonymity, though these are not included in this rating.
%Partial support indicates that the sender's IP address is protected and the sender's email address and mail server are disposable and open access, respectively, largely mooting the concern of leaking this information.

\col{S}{phishing}{Easy to detect phishing}
\subparagraph{\ref{col:phishing} \ref{col-text:phishing}}
Full support indicates that it is easy for a user to identify when an email is a phishing email.
Partial support indicates that easy identification of phishing email is only possible for long-standing contacts---for example, using key continuity~\cite{garfinkel2005johnny}.

\subsubsection{Utility Properties (\ref{col:provider-choice}--\ref{col:persistent-access})}\hfill

\col{T}{provider-choice}{Supports user choice of email providers}
\subparagraph{\ref{col:provider-choice} \ref{col-text:provider-choice}}
Full Support indicates that a user can use an email provider of their choice.
A rating of no support indicates that the system is dependent on a particular provider or employer.

\col{T}{server-choice}{Supports user choice of identity provider}
\subparagraph{\ref{col:server-choice} \ref{col-text:server-choice}}
An identity provider is the party that is responsible for binding a person's subject name or email address to their public key, such as a Certification Authority or trusted key server.
Full Support indicates that a user can use an identity provider of their choice.
%A rating of no support indicates that the system assumes a centralized identity provider.

\col{T}{content-processing}{Supports server-side content processing}
\subparagraph{\ref{col:content-processing} \ref{col-text:content-processing}}
Full support indicates server-side content processing is possible, for example to provide spam and malware filtering, to identify high priority emails, or to automatically label or reply to messages.
Note that research into computing on encrypted data is active and promises to enable the composition of message confidentiality and content processing~\cite{Kam15}, however these techniques have not yet been applied to secure email.
%we assume in our evaluation that only existing techniques are utilized.
%It is possible to use existing multi-party computation techniques to allow a mail server and an escrow server to provide privacy-preserving content processing, but no systems currently leverage this approach and so no end-to-end encrypted email system receives full support for this property.

\col{T}{persistent-access}{Provides persistent access to email}
\subparagraph{\ref{col:persistent-access} \ref{col-text:persistent-access}}
%Users have come to expect permanent access to their email messages and many consider it unacceptable to lose that access.
%For key management schemes (Table~\ref{tab:taxonomy-key-management}), full support is awarded if a user is able to recover their lost private keys without the need to remember or store a secret value (e.g., a password or a printed code); partial support indicates that private key recovery is possible, but only if the user remembers or has access to a secret value.
Full support indicates the user has persistent access to their email---whether through private key recovery or some other mechanism---without the need to remember or store a secret value.
Partial support indicates that persistent access to their email is possible, but only if the user remembers or has access to a secret value. We assume that access to the private key is what is important here---the actual emails can be stored in encrypted form in the cloud or on a local hard drive and backed up for resilience to hard drive failure.

% \col{T}{interop}{Supports interoperability}
% \item[\ref{col:interop}] \ref{col-text:interop}.
% Plaintext email systems generally allows users to send and receive email with all other email users, regardless of the system they use. Likewise, users who have adopted a secure email system may expect that they can email any other user securely who has another kind of secure email system, without having to worry about any details of how the systems interoperate.  Because systems have varying trust models, we assume that interoperability of secure email systems has a goal of protecting users from passive attacks only. We award full support if a system enables interoperable secure email with at least one other system. We also award full support for systems using plaintext email because using plaintext is a choice made to support interoperability, at the expense of numerous security properties. No support is given if a system is a walled garden, supporting only users of that exact system.

\subsubsection{Deployability Properties (\ref{col:email-client}--\ref{col:infrastructure})}\hfill

\col{D}{email-client}{No client software updates needed}
\subparagraph{\ref{col:email-client} \ref{col-text:email-client}}
Full support indicates that there is no need to update existing email clients or adopt a new email client.
As the need to change---not the magnitude of the change---is the main hindrance, there is no partial support for this or the following two properties.
% We do not award partial support, highlighting that the requirement to change is the main hindrance, not the magnitude of the change.
%Requiring installation of a browser extension earns a rating of no support.

\col{D}{email-server}{No email server updates needed}
\subparagraph{\ref{col:email-server} \ref{col-text:email-server}}
Full support indicates that there is no need to update existing email servers or adopt new email servers in order to support the secure email system.
% For the same reason as in \ref{col:email-client}, we do not award partial support.

\col{D}{infrastructure}{No infrastructure updates needed}
\subparagraph{\ref{col:infrastructure} \ref{col-text:infrastructure}}
Full support indicates that there is no need to update existing non-email infrastructure or adopt new non-email infrastructure.
% For the same reason as in \ref{col:email-client}, we do not award partial support.

\subsubsection{Usability Properties (\ref{col:key-discovery}--\ref{col:key-validation})}\hfill

% \col{U}{non-participants}{Easy to send to non-participants}
% \item[\ref{col:non-participants}] \ref{col-text:non-participants}.
% Reducing the number of steps an individual needs to complete before receiving an encrypted email for the first time makes it easier to adopt secure email.
% Full Support indicates that a recipient can receive encrypted email without needing to take any action (\eg installing software, generating a key pair).
% We award partial support if the only action a recipient needs to take is installing a software package with minimal or no configuration required---for example, entering their email credentials.
% A rating of no support indicates that the configuration is significant or it requires interaction with another user.

%\col{U}{key-discovery}{Effortless same system encryption key discovery}
\col{U}{key-discovery}{Effortless encryption key discovery}
\subparagraph{\ref{col:key-discovery} \ref{col-text:key-discovery}}
%We rate systems based on whether discovery of the recipient's encryption key is effortless for users of the same system, meaning the same deployment of a walled garden or the same centralized key server.
%For open systems we define the same system to encompass any software that implements the system.
Full support indicates that an email client can automatically acquire any recipient’s encryption key who is using the same system (e.g., secure mailbox provider, corporate email server, or public key directory).
Partial support indicates the system distributes encryption keys by automatically attaching them to outgoing emails.
This requires a sender to first receive email from a recipient before they can send encrypted email to that contact.

%   In order to send an encrypted email, it is necessary to first retrieve the recipient's public key.\footnote{Note, that if a public key is used for signing it is trivial to discover it as that public key can be attached to the signed email.}
% Full support indicates that an online user can automatically acquire any other user's public key who is using the same system.
% For systems (Table~\ref{tab:taxonomy-systems}), we also award partial support if effortless public key discovery is available, but only for users with the same configuration (e.g., same trusted introducers, key server, or email provider).
% For key management schemes (Table~\ref{tab:taxonomy-key-management}), partial support is not applicable as it represents an implementation decision to not support interoperability and not a fundamental limitation of the key management scheme.
% For example, any current system rating partial support could achieve full support by moving to a single global key server or providing a global mechanism for identifying which key server is used for which users.

\col{U}{key-validation}{Effortless encryption/signing key validation}
\subparagraph{\ref{col:key-validation} \ref{col-text:key-validation}}
Full support indicates a system automates validation, with a public key audit trail (\ref{col:public-key-audit}) and responsive public key revocation (\ref{col:revocation}).
Partial support indicates a system automates validation using a trusted key server or Certification Authority, with manual key validation needed only when there is concern that the the trusted entity might be acting maliciously.

\subsection{Priority Rankings}

% !TEX root = ../main.tex

\begin{table}[t]
	\setuptable
	\setlength{\arrayrulewidth}{1pt}
	
	\begin{tabular}{ L | *{8}{C} | *{4}{C} | *{3}{C} | *{2}{C} |}
		&
		\headtilt{\ref{col-full:confidentiality}} &
		\headtilt{\ref{col-full:integrity}} &
		\headtilt{\ref{col-full:private-key}} &
		\headtilt{\ref{col-full:exceptional-access}} &
		\headtilt{\ref{col-full:revocation}} &
		\headtilt{\ref{col-full:public-key-audit}} &
		\headtilt{\ref{col-full:privacy}} &
		\headtilt{\ref{col-full:phishing}} &

	    \headtilt{\ref{col-full:provider-choice}} &
		\headtilt{\ref{col-full:server-choice}} &
		\headtilt{\ref{col-full:content-processing}} &
		\headtilt{\ref{col-full:persistent-access}} &

		\headtilt{\ref{col-full:email-client}} &
		\headtilt{\ref{col-full:email-server}} &
		\headtilt{\ref{col-full:infrastructure}} &

		\headtilt{\ref{col-full:key-discovery}} &
		\headtilt{\ref{col-full:key-validation}} \\
		\cline{2-18}

		\textit{Stakeholder} &
		\multicolumn{8}{c|}{\textit{Security}} &
		\multicolumn{4}{c|}{\textit{Utility}} &
		\multicolumn{3}{c|}{\textit{Deploy.}} &
		\multicolumn{2}{c|}{\textit{Usab.}}\\

		%%%%%%%%%%%%%%%%%%%%%%%%%%%%%%%%%%%%%%%%%%%%%%%%%%%%%%%%%%%%%%%%%%%%%%%%%%%%%%%%%%%%
		\hline

		Enforcement
		&\na &\na	&\na &\anti	&\na &\na	&\anti &\na
		&\na &\na &\na &\na
		&\na &\na &\na
		&\na &\na	\\

		%%%%%%%%%%%%%%%%%%%%%%%%%%%%%%%%%%%%%%%%%%%%%%%%%%%%%%%%%%%%%%%%%%%%%%%%%%%%%%%%%%%%

		\hline

		Email Service Providers
		&\highprt	&\highprt	&\na &\na	&\na &\na	&\na &\high
		&\low &\low &\high &\high
		&\high &\tussle &\high
    	&\na &\na	\\

		Typical Users
		&\highprt	&\highprt	&\na &\tussle	&\na &\na	&\na &\high
    	&\high &\low &\high &\high
		&\high &\na &\na
	  	&\na &\na	\\

		Enterprise Organizations
		&\highprt	&\highprt	&\na &\anti	&\high &\high &\na	&\high
		&\low &\low &\high &\high
		&\low	&\low	&\high
		&\high &\high	\\

		%%%%%%%%%%%%%%%%%%%%%%%%%%%%%%%%%%%%%%%%%%%%%%%%%%%%%%%%%%%%%%%%%%%%%%%%%%%%%%%%%%%%
		\hline

		Secure Mailbox Providers
		&\high &\high &\highprt &\high &\high	&\low &\na &\high
		&\low	&\low &\low &\low
		&\na &\na &\na
		&\high &\high	\\

		Privacy Enthusiasts
		&\high &\high	&\tussle &\high	&\high &\low &\low &\high
		&\high &\high &\low	&\tussle
		&\low	&\na &\na
		&\high &\high	\\

		Vulnerable Users
		&\high &\high	&\high &\high	&\high &\high &\high &\high
		&\high &\high &\anti &\anti
		&\na &\na	&\na
		&\low	&\low	\\

		%%%%%%%%%%%%%%%%%%%%%%%%%%%%%%%%%%%%%%%%%%%%%%%%%%%%%%%%%%%%%%%%%%%%%%%%%%%%%%%%%%%%
		\hline

	\end{tabular}
	
	\hfill\\
	\begin{tabular}{Lp{2em}L}
          \high~high priority for full support & &
          \highprt~high priority for partial support \\
          \low~low priority & &
          blank means a non-priority or not applicable
	\end{tabular}

	\begin{tabular}{c}
		\tussle~there is disagreement within the stakeholder group about the priority of this property \\
        \anti~high priority for no support
	\end{tabular}
	
	\vspace{1\baselineskip}
	
	\caption{Stakeholder priorities.\label{tab:priorities}}

\end{table}

% JC: Priorities 

Table~\ref{tab:priorities} ranks the priority given by stakeholders to each property. A given priority can be a high, low, or a non-priority. In some cases, we rate a stakeholder as highly valuing partial support of a property. We also identify several cases where a stakeholder has a high priority that the property is \textit{not} met, meaning it is antithetical to their goals. 

We lightly clustered the stakeholders into three groups. Enforcement has unique priorities for the targets of their investigation; priorities are to backdoor completely confidential and anonymous communication. The second cluster generally prioritizes utility and deployability, while the third prefers security. We accept that the reader may disagree with some rankings but believe the framework enables a useful discussion of tradeoffs that are often otherwise glossed over.

% JC: Tussles defined and illustrated

There are several cases where we found disagreement within a stakeholder group regarding the priority of a given property (marked \tussletext). An example is preventing exceptional access to email (\ref{col:exceptional-access})---typical email users are divided between those who advocate for government surveillance of email and who are willing to accept government access to email on presentation of a warrant, and those who strongly prefer end-to-end encryption that would prevent exceptional access. Likewise, privacy enthusiasts are split on whether there is a high priority on ensuring that private keys are accessible only to users (\ref{col:private-key}), with a minority placing a high priority on this property but others accepting password-protected cloud storage of a private key. Privacy enthusiasts are also split on whether persistent access to email is a high priority (\ref{col:persistent-access}), along similar lines. Finally, while many email service providers place a high priority on not being required to deploy new email-related servers to support a given technology (\ref{col:email-server}), this is likely not a high priority for larger providers. For example, large providers have shown a willingness to adopt best practices such as STARTTLS and DKIM more rapidly.

% JC: Partial support defined and illustrated

In several cases, stakeholders have a high priority for partial support of a property but do not want it fully (or universally) supported (marked \highprttext). All stakeholders, aside from enforcement, prefer that emails are protected from eavesdropping by third parties (\ref{col:confidentiality}). However certain stakeholders want read capabilities for some email. For example, an enterprise may want to run automated services on their employees' plaintext emails---for security, compliance or other reasons---but do not want the emails accessible in plaintext by anyone outside of the enterprise, or even anyone within the enterprise that is not a party to the email. Similarly, enterprises and service providers may want the ability to modify email messages (S2) to protect their users (remove malware or insert a phishing warning) without disrupting message authentication. Users may want this protection as well.  

As a final example of partial support, secure mailbox providers offer users the ability to control their own signing and encryption keys (\ref{col:private-key}) but balance this with some usability features. For example, storing password-protected decryption keys in the cloud allows users to check their email from new devices without transferring their keys, while it limits the provider's access to their users' decryption keys. This is in contrast to a (normal) email service provider that, if it supported encryption and signatures at all, would give customers the additional usability feature of backing up their private decryption keys, enabling key recovery and the ability to read past encrypted emails. Note that private keys for signing do not require backup as users can generate new ones, although the old public signature keys should be maintained for verification of past emails (or revoked if the signing key is stolen as opposed to lost).

\subsubsection{Inter-Stakeholder Tussles}
Table~\ref{tab:priorities} illustrates the reality that there are significant disagreements between stakeholders in the secure email space and that no single solution will satisfy them all. The strongest disagreements happen in columns where at least one stakeholder fully supports a property (marked \hightext) while another strongly opposes it it (marked \antitext). The four high conflict properties are exceptional access (\ref{col:exceptional-access}), sender pseudonymity (\ref{col:privacy}), server-side content-processing (\ref{col:content-processing}), and persistent access (\ref{col:persistent-access}). 

The conflict between enforcement and other stakeholders over exceptional access (\ref{col:exceptional-access}) and sender pseudonymity (\ref{col:privacy}) is well-known in both secure email and other technical domains: web browsing, network traffic, server IP addresses and locations, and payment systems. We emphasize again that the enforcement stakeholder category captures enforcement's preferences for the targets of their investigations and actions, while the agents themselves are better aligned with privacy enthusiasts, and agents could use (or create) vulnerable users through their investigations.

High conflict also exists over server-side content-processing (\ref{col:content-processing}) for spam, malware filtering, classification, or automatic replies; and persistent access (\ref{col:persistent-access}) which indicates that the user can recover their access and archive after losing their authentication credentials. This conflict illustrates an important result: some of the most fundamental disagreements occur over the utility properties of a secure email system. Email service providers, typical users, and enterprise organizations all place a high value on content processing and persistent access. Yet, these are mostly low priorities for the other stakeholders and, in some cases, antithetical to the principles held by vulnerable users who prioritize exclusive access to their email with no backdoors. Even if it means managing a secret value that only they know, they accept the risk of key loss being permanent. 

% JC: General comments

The tussles among stakeholders help explain the history of how this space has evolved. The needs of typical users are largely met by email service providers; these two stakeholders disagree mainly on deployment properties that affect only the service provider (\ref{col:email-server}, \ref{col:infrastructure}), along with a tussle over exceptional access (\ref{col:exceptional-access}). Privacy enthusiasts have a demonstrated history of highly valuing end-to-end encryption (hence the development of PGP and person-to-person key exchange), but it is not a priority for email service providers and typical users, and this explains why it is not pursued more broadly.  The needs of some enterprise organizations to deploy secure email explains why they often adopt S/MIME based products. They need encryption within the organization, plus escrow of private keys and content processing.  They also have the IT budget to provide a seamless user experience. 

Privacy enthusiasts overlap significantly with enterprise organizations, but disagreements on private key storage (\ref{col:private-key}), server-side content processing (\ref{col:content-processing}) and persistent access (\ref{col:persistent-access}) make finding common ground difficult. Privacy enthusiasts also overlap with vulnerable users but vulnerable users will tolerate poor usability and a lack of features to maximize security. To our knowledge, no major commercial provider currently meets the needs of vulnerable users.

Most email service providers prioritize opportunistic encryption with TLS.  Secure email providers have emerged, with priorities that mostly match those of privacy enthusiasts, some of whom may previously have used PGP-based services. Some privacy enthusiasts would prefer the private key is only accessible to themselves (\ref{col:private-key}), but due to the loss of grass-roots support for PGP, the only apparent feasible alternative is password-protected keys used in secure webmail. The services offered by secure email providers have supported vastly more users of secure email than PGP ever did. However their business model naturally means some deployment properties cannot be met, hence requiring users to use new email software.

\section{Technical Evaluation of Secure Email Approaches}
\label{sec:evaluations}

In the previous section, we identified how stake holders prioritize various secure email properties.
In this section we turn our attention to the building blocks for secure email---cryptographic primitives, key management schemes, and system designs---and their support for these properties.
The list of building blocks in this section are pulled from our review of the secure email systems and literature (\S\ref{sec:stakeholders}) and is not intended to be exhaustive, but rather representative, focusing on the most commonly used and discussed high-level approaches.
Our evaluation of these building blocks helps explore how well stakeholders are served by existing approaches and to identify common limitations of these approaches.

\subsection{Cryptographic Primitives}
\label{sec:evaluations-crypto}

\newcounter{CryptoPrimitiveCounter}
\makeatletter
\newcommand*{\crpto}[2]{%
	\refstepcounter{CryptoPrimitiveCounter}\def\@currentlabel{C\arabic{CryptoPrimitiveCounter}}\label{cp:#1}%
	\def\@currentlabel{#2}\label{cp-text:#1}%
	\ref{cp:#1} & \ref{cp-text:#1}%
}
\makeatother

\begin{table}
	\setuptable

	\begin{tabular}{ R | LL | L | *{3}{C} | *{1}{C} | *{2}{C} |}
		\head{R}{} & \head{c}{\#} & \head{L}{Cryptographic Approach} & Reference &

		\headtilt{\ref{col-full:confidentiality}} &
		\headtilt{\ref{col-full:integrity}} &
		\headtilt{\ref{col-full:privacy}} &
		
		\headtilt{\ref{col-full:content-processing}} &
		
		\headtilt{\ref{col-full:email-client}} &
		\headtilt{\ref{col-full:email-server}} \\
		\hline

		\multirow{6}{*}{MTA-based}
%		\rotatebox{90}{MTA-based}}

		&\crpto{plaintext}{Plaintext email} & \S\ref{sec:baseline}
		&\none		&\none		&\none
		&\full		&\full		&\full		\\

		&\crpto{link-enc}{Link encryption} & \S\ref{sec:link-encryption}
		&\none		&\none		&\none
		&\full		&\full		&\none		\\

		&\crpto{mta-enc}{Provider-to-provider encryption} & \S\ref{sec:hosted}
		&\prt		&\none		&\none
		&\full		&\full		&\none		\\

		&\crpto{mta-sig}{Provider-to-provider signing} & \S\ref{sec:domain-auth}
		&\none		&\prt		&\none
		&\full		&\full		&\none		\\

		&\crpto{layered}{Layered encryption} & \S\ref{sec:layered}
		&\prt		&\none		&\full
		&\full		&\none		&\none		\\

		\hline
        \multirow{2}{*}{End-to-End}
      % \rotatebox{90}{E2E}

		&\crpto{e2e-enc}{End-to-end encryption} & \S\ref{sec:smime}, \S\ref{sec:pgp}
		&\full		&\none		&\none
		&\none		&\none		&\full		\\

		&\crpto{e2e-sig}{End-to-end signing} & \S\ref{sec:smime}, \S\ref{sec:pgp}
		&\none		&\full		&\none
		&\full		&\none		&\full		\\

		\hline
	\end{tabular}

	\hfill\\
	\fullsymbol~full support, \prtsymbol~partial support, \nonesymbol~no support, blank means not applicable

	\hfill\\
	\caption{Comparative evaluation of cryptographic approaches used to enhance email security.\label{tab:taxonomy-crypto}}
\end{table}

We identified seven common cryptographic primitives used in secure email.

\subparagraph{\ref{cp:plaintext} \textit{\ref{cp-text:plaintext}}}
Email without any cryptographic protections is used as a baseline for comparison to the other primitives.
	
\subparagraph{\ref{cp:link-enc} \textit{\ref{cp-text:link-enc}}}
Encrypting the email content during transmission (\eg TLS) prevents passive attackers from reading email messages.
Protection against an active man-in-the-middle attacks depends on which key management primitives are composed with this primitive.
Note that this primitive only offers complete protection if all links along the transmission path are properly encrypted.
	
\subparagraph{\ref{cp:mta-enc} \textit{\ref{cp-text:mta-enc}} and \ref{cp:mta-sig} \textit{\ref{cp-text:mta-sig}}}
The sender's mail provider encrypts or signs the body of email messages, which is then decrypted or verified by the recipients' email providers.
This improves upon link encryption (\ref{cp:link-enc}) by removing the ability of intermediate MTAs to see or modify messages.
	
\subparagraph{\ref{cp:layered} \textit{\ref{cp-text:layered}}}
Layered encryption---alternatively referred to as onion encryption---encrypts email messages (the body and parts of the envelope) in layers, with one layer being removed at each MTA. Onion encryption for email has been proposed in the Dark Internet Mail Environment (DIME)~\cite{dime}.
Encryption is performed by the sender, and then the encrypted message is sent to the sender's mail provider along with the address for the MTA that can unwrap the first layer of encryption.
When received by that MTA, one layer of encryption will be removed revealing the next MTA for the message to be forwarded to, with this pattern continuing until received by the recipient's mail provider.
As long as this chain---as selected by the sender---contains at least one intermediate MTA, neither the sender's or recipients' email providers will learn who received or sent, respectively, the email.
In practice, this scheme would be used with end-to-end encryption (\ref{cp:e2e-enc}) and signing (\ref{cp:e2e-sig}).
	
\subparagraph{\ref{cp:e2e-enc} \textit{\ref{cp-text:e2e-enc}} and \ref{cp:e2e-sig} \textit{\ref{cp-text:e2e-sig}}}
Email is encrypted and signed with cryptographic keys associated with end-users.
Cryptographic operations are executed at the end-user (client) devices.
	
\subsubsection{Evaluation:}

Table~\ref{tab:taxonomy-crypto} evaluates how cryptographic primitives for encryption and digital signing meet relevant security, deployment, and usability properties.
%In this and the next section we limit the properties to a relevant subset in order to simplify our presentation.
We divide the cryptographic primitives into those that are MTA-based versus those that operate end-to-end.
Justification for the ratings is given in \S\ref{appx:crypto}.

We draw several conclusions from this analysis. First, link encryption does not get full support on any of the security properties since any MTA that is on a delivery path can access and modify plaintext email. Referring back to Table~\ref{tab:priorities}, this is good enough for email providers and for typical users, provided it is implemented more widely. Second, end-to-end encryption breaks server-side content processing (\ref{col:content-processing}), which makes it a non-starter for some stakeholders. Third, the MTA-based approaches satisfy the deployment properties and the usability property.  Yet, we know from measurement studies ~\cite{durumeric2015neither,foster2015security} that many smaller email providers have not yet adopted this technology. It is not clear whether this lack of adoption is due to lack of interest or if the technologies are difficult to use. If the latter, then this may be low-hanging fruit that could be improved with some attention to the usability of available tools.

\subsection{Key Management Schemes}
\label{sec:evaluations-keymanagement}

Key management schemes are responsible for verifying encryption and signing certificates as well as allowing users to discover other users' public keys.
We identified ten common key management schemes used in secure email.

\subparagraph{\ref{km:ca} \textit{\ref{km-text:ca}}}
After generating a key pair, a user submits their public key and distinguished name (\ie email address) to a Certification Authority.
After validating the user's ownership of the distinguished name and the private key associated with the submitted public key, the Certification Authority issues a signed certificate that binds the public key to the distinguished name.
When receiving a certificate,
%---either through an out-of-band communication channel or using a certificate directory (\ref{km:ca-directory})---
a relying client validates the certificate, including verifying the signature, with trust anchored by a public key in the system's trusted root store.

\subparagraph{\ref{km:ca-directory} \textit{\ref{km-text:ca-directory}}}
Extends \ref{km:ca} by adding a third party that stores and disseminates
certificates.
Disseminated certificates are verifiable because an authoritative third party (\ie a Certification Authority) signed them.

% \subparagraph{\ref{km:ca-escrow} \textit{\ref{km-text:ca-escrow}}}
% This scheme extends \ref{km:ca-directory} with storage of users private keys.
% Optionally, instead of having key pairs generated and uploaded by users, the key escrow server can generate and store keys as they are needed.
% Any user can be sent an encrypted email even if they haven't generated a key pair as the key escrow server can generate and distribute the key for the users, with the user only retrieving their key pair after they have received their first encrypted email.

\subparagraph{\ref{km:mkd} \textit{\ref{km-text:mkd}}}
Users are responsible for exchanging and validating public keys amongst themselves.
This may happen in person---for example, at a PGP key signing party.
Alternatively, a user could retrieve a key from a public source (\eg a personal website), then validate that key by contacting the other user over an out-of-band channel (\eg phone call).
%Users cannot send encrypted email to or verify signatures from users they have no who they have not yet exchanged keys with.

\subparagraph{\ref{km:mkd-web} \textit{\ref{km-text:mkd-web}}}
This scheme builds on \ref{km:mkd} by adding trusted introducers (\eg as in PGP).
In addition to accepting certificates that they have personally exchanged, relying clients will also accept as valid certificates that have been signed (endorsed) by one or more trusted introducers. Trusted introducers may be
seen as bestowing CA-like powers to non-expert users.
%In addition to accepting certificates that they have personally exchanged, users will also accept certificates that have been signed by a sufficient number of endorsers.
This scheme does not aid in certificate discovery for encryption.

\subparagraph{\ref{km:mkd-server} \textit{\ref{km-text:mkd-server}}}
An extension of \ref{km:mkd-web} that adds a third-party server responsible for publishing public keys of users, including metadata regarding the owner of the key (such as name and email address) and signatures attesting to the binding of the key with the owner (\ie trusted introducer signatures). Clients are still responsible for determining the validity of any signatures on the packets and deciding if they are trustworthy.

\subparagraph{\ref{km:trusted-server} \textit{\ref{km-text:trusted-server}}}
A server that publishes public keys of users, asserting a binding of the key to an owner (such as an email address). The server is trusted to provide valid bindings.  This scheme relies on a secure connection between clients and the public key server---for example, using TLS---rather than using certificates.

\newcounter{KeyManagementCounter}
\makeatletter
\newcommand*{\keyman}[2]{%
	\refstepcounter{KeyManagementCounter}\def\@currentlabel{KM\arabic{KeyManagementCounter}}\label{km:#1}%
	\def\@currentlabel{#2}\label{km-text:#1}%
	\ref{km:#1} & \ref{km-text:#1}%
}
\makeatother

\begin{table}
	\setuptable

	\begin{tabular}{ LL | L | *{3}{C} | *{1}{C} | *{1}{C} | *{2}{C} |}
		\# & \head{L}{Key Management Scheme} & \head{c|}{Ref.} &

		\headtilt{\ref{col-full:private-key}} &
		\headtilt{\ref{col-full:revocation}} &
		\headtilt{\ref{col-full:public-key-audit}} &
		
		\headtilt{\ref{col-full:persistent-access}} &

		\headtilt{\ref{col-full:infrastructure}} &

		\headtilt{\ref{col-full:key-discovery}} &
		\headtilt{\ref{col-full:key-validation}} \\
		\hline

		\keyman{ca}{Certificate authority} & \S\ref{sec:link-encryption}
		&\full		&\full		&\none
		&\none		&\full
		&\prt	  	&\full		\\

		\keyman{ca-directory}{Certificate directory} & \S\ref{sec:smime}
		&\full		&\full 		&\none
		&\none		&\none
		&\full		&\full		\\

%		\keyman{ca-escrow}{Directory + key escrow} & \S\ref{sec:smime}
%		&\none		&\none		&\full		&\full		&\none
%		&\full		&\full		\\

		\hline

		\keyman{mkd}{Manual key distribution} & \S\ref{sec:pgp}
		&\full		&\none		&\none		&\none		&\full
		&\none		&\none		\\

		\keyman{mkd-web}{Web of trust (WoT)} & \S\ref{sec:pgp}
		&\full		&\none		&\none		&\none		&\full
		&\none		&\none		\\

		\keyman{mkd-server}{WoT + key packet server} & \S\ref{sec:pgp}
		&\full		&\none		&\none		&\none		&\none
		&\full		&\none		\\

		\hline
		\keyman{trusted-server}{Trusted public key server} & \S\ref{sec:trusted}
		&\full		&\full	&\none	&\none		&\none
		&\full		&\prt		\\

		\keyman{trusted-ledger}{Audited public key server} & \S\ref{sec:audited}
		&\full		&\full		&\full		&\none		&\none
		&\full		&\full		\\

		\hline

		\keyman{tofu}{Trust on first use} & \S\ref{sec:tofu}
		&\full		&\none		&\none		&\none		&\full
		&\prt		&\none		\\

		\keyman{shared-secrets}{Shared secrets} & \S\ref{sec:webmail}
		&\none		&\none	&\na			&\none		&\full
		&\none		&\na		\\

	  	\hline
	  	
	  	\keyman{ibe}{Identity-based encryption} & \S\ref{sec:ibe}
	  	&\none		&\none		&\none		&\full		&\none
	  	&\full		&\full		\\

		\keyman{escrow}{Key escrow} & \S\ref{sec:ibe}
		&\none		&\na		&\na		&\full		&\none
		&\na		&\na		\\

		\hline
	\end{tabular}

	\hfill\\
	\fullsymbol~full support, \prtsymbol~partial support, \nonesymbol~no support, blank means not applicable

	\hfill\\
	\caption{Comparative evaluation of key management schemes used to enhance email security.\label{tab:taxonomy-key-management}}
\end{table}

\subparagraph{\ref{km:trusted-ledger} \textit{\ref{km-text:trusted-ledger}}}
This scheme extends \ref{km:ca-directory} or \ref{km:trusted-server} by allowing the operation of the key server to be audited by external entities.
In particular, auditors can examine a history of all certificates or key packets that the ledger has made available for any entity, allowing them to detect rogue certificates or keys and server equivocation.

\subparagraph{\ref{km:tofu} \textit{\ref{km-text:tofu}}}
With this scheme, users attach their public keys or certificates to all outgoing email messages, regardless of whether the messages are signed or encrypted.
When a recipient receives a public key or certificate for the first time, the recipient's client associates that public key or certificate with the sender's identity.
In the future, if a new public key or certificate is received, the recipient's client must decide whether it is a valid update or an attack. In secure messaging applications this is done with out-of-band key verification, known as an authentication ceremony, but this leads to several usability challenges~\cite{vaziripour2017you}.

\subparagraph{\ref{km:shared-secrets} \textit{\ref{km-text:shared-secrets}}}
Instead of using key pairs (\ie public key cryptography), clients can encrypt messages using a shared secret (\ie symmetric key cryptography).
These secrets could be a symmetric key or more likely a string both users know (\eg password) that is hashed to generate a symmetric key.
Secrets are shared in person or over an out-of-band channel that provides confidentiality.

\subparagraph{\ref{km:ibe} \textit{\ref{km-text:ibe}}}
Identity-based encryption~\cite{shamir1984identity} centers around the use of an IBE key server that stores a secret value. This secret value is used to generate a set of public parameters that are published by the IBE key server. Other clients can use these parameters to derive a public key for any email address.
The IBE key server uses its secret value to generate private keys for those public keys derived using the published public parameters.
When users wish to retrieve their private key, they authenticate to the IBE key server which then returns their private key.

\subparagraph{\ref{km:escrow} \textit{\ref{km-text:escrow}}}
This scheme provides a backup copy of private keys or storage of symmetric encryption keys.
This backup can be used to restore access to these keys if lost by users.
%Some organizations may operate a key escrow server that generates and stores keys on demand, so that a user can be sent an encrypted email even if they haven't generated a key pair yet.
%The escrow server in this case needs a method to identify users, for example using an enterprise authentication system.

\subsubsection{Evaluation:}

Table~\ref{tab:taxonomy-key-management} evaluates how key management schemes meet relevant security, deployment, and usability properties. Justifications for the ratings are in \S\ref{appx:keymanagement}.

We note that most systems requiring no new non-email infrastructure (\ref{col:infrastructure}) have poor usability properties. This illustrates one reason why certificate authorities, manual key distribution, and trust on first use have not been widely used for secure email, and why shared secrets have seen some use among secure email providers. Two schemes, certificate directories with private key escrow and identity-based encryption---meet all usability properties. Stakeholders who highly value persistent access (\ref{col:persistent-access}), such as enterprises, have primarily adopted the former.  Likewise, trusted public key servers offer a combination of security and usability that matches what secure email providers are offering. Adding monitors to the public key servers would add a public key audit trail (\ref{col:public-key-audit}) and thus offer some additional promise for stakeholders valuing this property (e.g. privacy-essential users).

\subsection{Secure Email Designs}
\label{sec:evaluations-systems}

% !TEX root = ../main.tex

\newcounter{SystemCounter}
\makeatletter
\newcommand*{\sys}[2]{%
	\refstepcounter{SystemCounter}\def\@currentlabel{SYS\arabic{SystemCounter}}\label{sys:#1}%
	\def\@currentlabel{#2}\label{sys-text:#1}%
	\ref{sys:#1} & \ref{sys-text:#1}%
}
\makeatother

\newcommand*{\rowheader}[2]{\multirow{#1}{1.8cm}{#2}}

\ifuselandscape{\begin{landscape}\begin{table}}{\begin{sidewaystable}\vspace{5.7in}}
	\setuptable
	
	\begin{adjustbox}{max width=\ifuselandscape{7.5in}{\textwidth}}
	\begin{tabular}{ L | LL | L | L |  *{8}{C} | *{4}{C} | *{3}{C} | *{2}{C} |}
		\multicolumn{5}{L|}{} &
		
		\headtilt{\ref{col-full:confidentiality}} &
		\headtilt{\ref{col-full:integrity}} &
		\headtilt{\ref{col-full:private-key}} &
		\headtilt{\ref{col-full:exceptional-access}} &
		\headtilt{\ref{col-full:revocation}} &
		\headtilt{\ref{col-full:public-key-audit}} &
		\headtilt{\ref{col-full:privacy}} &
		\headtilt{\ref{col-full:phishing}} &
		
		\headtilt{\ref{col-full:provider-choice}} &
		\headtilt{\ref{col-full:server-choice}} &
		\headtilt{\ref{col-full:content-processing}} &
		\headtilt{\ref{col-full:persistent-access}} &
		
		\headtilt{\ref{col-full:email-client}} &
		\headtilt{\ref{col-full:email-server}} &
		\headtilt{\ref{col-full:infrastructure}} &
		
		\headtilt{\ref{col-full:key-discovery}} &
		\headtilt{\ref{col-full:key-validation}} \\
		\cline{6-22}
		
		Family & \# & System & Components & Format &
		\multicolumn{8}{c|}{\textit{Security}} &
		\multicolumn{4}{c|}{\textit{Utility}} &
		\multicolumn{3}{c|}{\textit{Deploy.}} &
		\multicolumn{2}{c|}{\textit{Usab.}} \\
		\hline
		
		\rowheader{3}{MTA-based}
		& \sys{plaintext}{Baseline email} & \ref{cp:plaintext} & Plaintext 
		& \none & \none & \na & \none & \na & \na & \none & \none 
		& \full & \na & \full & \full 
		& \full & \full & \full 
		& \na & \na \\

		& \sys{modern-email}{Email + TLS and DKIM} & \ref{cp:link-enc}, \ref{cp:mta-sig}, \ref{km:ca} & Plaintext 
		& \none & \prt & \na & \none  & \na   & \na   & \none & \none 
		& \full & \na & \full       & \full 
		& \full & \none & \none 
		& \na & \na \\

		& \sys{mixminion}{Mixminion Remailer}~\cite{Gol07} & \ref{cp:layered}, \ref{km:ca-directory} & Plaintext 
		& \prt & \none & \na & \none  & \na   & \na   & \full & \none 
		& \full & \na & \full       & \full 
		& \none & \none & \none 
		& \na & \na \\
		
		\hline
		\rowheader{3}{Walled Garden}
		
	   & \sys{smime}{Corporate S/MIME} & \ref{cp:e2e-enc}, \ref{cp:e2e-sig}, \ref{km:ca}, \ref{km:escrow} & S/MIME 
	   & \prt & \prt & \none       & \none  & \full & \none & \none & \none 
	   & \none & \none & \none       & \full 
	   & \none & \full & \none 
	   & \full & \prt \\
	
	   & \sys{hosted-smime}{Hosted S/MIME}~\cite{hostedsmime}     & \ref{cp:mta-enc}, \ref{cp:mta-sig}, \ref{km:ca}, \ref{km:escrow} & S/MIME 
	   & \prt & \prt & \none       & \none  & \full & \none & \none & \none 
	   & \none & \none & \full       & \full 
	   & \full & \none & \none 
	   & \prt & \prt \\
	
	   & \sys{secure-webmail}{ProtonMail}\cite{protonmail}       & \ref{cp:e2e-enc}, \ref{cp:e2e-sig}, \ref{km:trusted-server}, \ref{km:shared-secrets} & PGP 
	   & \full & \full & \prt & \prt   & \full & \none & \none & \none 
	   & \none & \none & \none       & \prt 
	   & \none & \full & \none 
	   & \full & \prt \\
		
		\hline
		\rowheader{4}{Open System}
		
		& \sys{pgp}{PGP} & \ref{cp:e2e-enc}, \ref{cp:e2e-sig}, \ref{km:mkd-server} & PGP 
		& \full & \full & \full       & \full  & \none & \none & \none & \none 
		& \full & \full & \none       & \none 
		& \none & \full & \none 
		& \none & \none \\

		& \sys{tofu}{Autocrypt}~\cite{autocryptspec} & \ref{cp:e2e-enc}, \ref{cp:e2e-sig}, \ref{km:tofu} & PGP 
		& \full & \full & \full       & \full  & \none & \none & \none & \none 
		& \full & \full & \none       & \none 
		& \none & \full & \full 
		& \prt & \none \\

		& \sys{johnny2}{Key continuity}~\cite{garfinkel2005johnny} & \ref{cp:e2e-enc}, \ref{cp:e2e-sig}, \ref{km:ca}, \ref{km:tofu} & S/MIME 
		& \full & \full & \full       & \full  & \none & \none & \none & \prt  
		& \full & \full & \none       & \none 
		& \none & \full & \full 
		& \prt & \none \\

		& \sys{ect}{Enhanced CT}~\cite{ryan2014enhanced} & \ref{cp:e2e-enc}, \ref{cp:e2e-sig}, \ref{km:trusted-ledger} & S/MIME 
		& \full & \full & \full       & \full  & \full & \full & \none & \none 
		& \full & \none & \none       & \none 
		& \none & \full & \none 
		& \full & \full \\
		
		\hline
		\rowheader{1}{Proprietary}
		
		& \sys{virtru}{Virtru}~\cite{virtru} & \ref{cp:e2e-enc}, \ref{cp:e2e-sig}, \ref{km:escrow} & Proprietary 
		& \full & \full & \none       & \prt   & \full & \none & \none & \none 
		& \full & \none & \none       & \full 
		& \none & \full & \none 
		& \full & \none \\
		
		%		&\sys{pwm}{Private Webmail~\cite{ruoti2016private}} &\ref{cp:e2e-enc}, \ref{cp:e2e-sig}, \ref{km:ibe} & Proprietary
		%		&\full		&\full &\none	&\prt	&\none	&\none		&\none		&\none
		%		&\full &\none  &\none		&\full
		%		&\none		&\full		&\none
		%		&\full		&\prt		\\
		
		\hline
	\end{tabular}
	\end{adjustbox}
	
	\hfill\\
	\fullsymbol~full support, \prtsymbol~partial support, \nonesymbol~no support, blank means not applicable

	\hfill\\	
	\caption{A comparative evaluation of system designs used to enhance email security.}
	\label{tab:taxonomy-systems}
\ifuselandscape{\end{table}\end{landscape}}{\end{sidewaystable}}

We finally turn our attention to secure email designs.
These designs incorporate the cryptographic primitives and key management schemes identified above, but make additional design decisions that further increase or limit the security, utility, deployability, or usability of the overall system design.
In this evaluation, we consider eleven system designs that are either widespread or are oft-cited as directions secure email should go.

\subparagraph{\ref{sys:plaintext} \textit{\ref{sys-text:plaintext}}}
This system design refers to sending and receiving plaintext email over unsecured links and is used as a baseline for comparison to the other designs.

\subparagraph{\ref{sys:modern-email} \textit{\ref{sys-text:modern-email}}}
This system design uses TLS where available to secure the links between users, email providers, and mail transfer agents (MTAs).
Additionally, Domain Keys Identified Email (DKIM) is used to sign email messages.

\subparagraph{\ref{sys:mixminion} \textit{\ref{sys-text:mixminion}}}
This system design uses layered encryption and mixminion relays between the sender's and recipients' mail providers.

\subparagraph{\ref{sys:smime} \textit{\ref{sys-text:smime}}}
This system design uses S/MIME to encrypt and sign messages.
Key pairs for each user are stored and distributed by an LDAP (or similar) server.
Public keys for a user are discoverable by other users of the enterprise or organization, but not by outside users.

\subparagraph{\ref{sys:hosted-smime} \textit{\ref{sys-text:hosted-smime}}}
This system design is a variant of S/MIME that has the email provider (in practice, Google) store each user's public and private keys to encrypt and decrypt messages for users.
This removes the need for email clients to support end-to-end encryption.

\subparagraph{\ref{sys:secure-webmail} \textit{\ref{sys-text:secure-webmail}}}
This is a secure webmail system design that provides automatic encryption among users of the system.
The email provider stores a user's password-protected private key so the user can access their email from multiple devices.
If senders wish to email recipients from other email providers, messages are encrypted using a shared secret.

\subparagraph{\ref{sys:pgp} \textit{\ref{sys-text:pgp}}}
This system design uses PGP, with keys distributed in person or using a key packet server.

\subparagraph{\ref{sys:tofu} \textit{\ref{sys-text:tofu}}}
This system design seeks to improve the adoption of end-to-end encrypted email. The goals of the design are to protect against passive eavesdropping, focus on incremental deployment, avoid asking users about keys, change only email clients, and use decentralized, in-band key discovery. Current specifications build on PGP and call for encryption keys to be exchanged automatically over email, with the first key received from a particular email address being bound to that email address (trust-on-first-use).
Key continuity could be used to help users identify malicious key changes, but has not currently been implemented.
% I'm not sure what this next sentence means. Cited the paper instead.
%As part of this scheme, we also include the work of Garfinkel and Miller on key continuity \cite{garfinkel2005johnny}.

\subparagraph{\ref{sys:johnny2} \textit{\ref{sys-text:johnny2}}}
This system design has users obtain self-signed certificates and then exchange them using trust-on-first-use.

\subparagraph{\ref{sys:ect} \textit{\ref{sys-text:ect}}}
This system design uses a Certification Authority and auditing via Certificate Transparency. A user's email provider typically acts as the identity provider, creating a certificate and uploading it to transparency logs.

\subparagraph{\ref{sys:virtru} \textit{\ref{sys-text:virtru}}}
This proprietary system design operates both a basic and advanced email encryption service. In basic operation, a client encrypts email using a symmetric key, then stores the key with a Virtru key escrow server. This server implements access control and delivers the symmetric key only to authorized recipients of an email. In the more advanced operation, a customer operates a trusted key server, which encrypts the symmetric key with a public key for the recipient. We rate Virtru with its basic mode because few details are publicly available about the key management used in the advanced mode.

%\subparagraph{\ref{sys:pwm} \textit{\ref{sys-text:pwm}}}
%This webmail system design uses key escrow based on identity-based encryption and a browser plugin that can be made compatible with the user's choice of email provider.

%\subparagraph{\ref{sys:confidante} \textit{\ref{sys-text:confidante}}}
%Extends \ref{sys:pgp} by replacing the manual key management, key packet server, and the web of trust with a fully automated key management scheme that relies on an auditable key server~\cite{melara2015coniks}.

\subsubsection{Evaluation:}

Table~\ref{tab:taxonomy-systems} evaluates secure email design paradigms composed from the cryptographic primitives and key management schemes. Justifications for the ratings are in \S\ref{appx:systems}. By comparing this table to Table~\ref{tab:priorities} we can identify designs that align well with the priorities of particular stakeholders. This comparison also helps illustrate how the lack of convergence to a single secure email system design is well explained by different stakeholder priorities.

In making this comparison, we note that unsecured email scores highly on deployability and relevant usability properties, meeting well the needs of typical email users and email providers. Adding best practices comes at a cost for providers, explaining why there is a long tail of providers not yet offering these services. Corporate S/MIME mostly matches the priorities of enterprises, but does require new infrastructure (\ref{col:infrastructure}) and new client software (\ref{col:email-client}), which could explain why it is not more widely used. Hosted S/MIME shifts the deployability responsibility toward providers and away from users, offering a system design for secure email for users that is similar to a corporate environment, provided users trust their email provider with their private key. PGP has significant usability problems, but secure webmail, offered by numerous secure email providers, aligns well with the needs of privacy enthusiasts, with the primary cost being some additional infrastructure and compromise on how private keys are stored. Adding automated key management would improve two security properties for secure webmail providers, while also adding interoperability, and thus represents an opportunity to better serve privacy enthusiasts. The relative low priority of interoperability, and the associated infrastructure costs, may explain why this hasn't happened yet.

Examining groups of properties yields additional insights. No system design fully meets all security properties, with metadata privacy and phishing prevention being significant gaps for most system designs. Deployability is clearly a hurdle for most system designs, particularly for those that offer additional security properties, such as a public key audit trail or responsive revocation. Usability offers a tussle among stakeholders. System designs ensuring the private key is accessible only to the user cannot provide content-filtering and usually cannot offer persistent access. However, if users are willing to give up these properties, the usability is generally not a significant hurdle since many system designs have adopted automated key management.

% !TEX root = ../main.tex

\section{Further Discussion}

After extensively reviewing the history of email, academic literature, discussing stakeholder priorities, and evaluating cryptographic primitives, key management schemes, and system designs, we highlight several critical points in understanding the state of secure email today.

\paragraph{A one-size-fits-all solution is unlikely.} 

It is clear from Table~\ref{tab:priorities} that stakeholders have conflicting priorities and that the needs of different stakeholders dictate diverging solutions. As such, it is unlikely that any single secure email system will be suitable for all users and their divergent use cases.  Furthermore, no single party controls the email ecosystem, and widespread deployment of secure email needs cooperation of numerous stakeholders. No one stakeholder has the capability to build (or the ability to demand) a secure email system that provides seamless interoperability for the billions of email users and supports email's many diverse uses. This means that even in the best case, with different solutions being adopted by different parties, there will almost surely be interoperability challenges that act as natural roadblocks and will require significant investment to overcome, if this is even possible.

\paragraph{The PGP web of trust remains unsuccessful after 25 years.}

The web of trust that is central to the original design of PGP---including manual key exchange and trusted introducers---has largely failed.  Its use is generally limited to isolated, small communities. Its appeal is that it allows quick, interoperable deployment in small groups without bureaucracy or costs of formal Certification Authorities, but in practice the downside is poor usability and lack of responsive revocation. Arguably, the resulting product indecision and non-interoperability has negatively impacted the deployment of secure email in general.

% Many long-term PGP supporters have abandoned the web of trust, and developers are pursuing alternative methods for distributing public keys.
% Secure webmail has supplanted traditional PGP clients. 
% As such, PGP has become more about the format of messages and keys, than the methods used to distribute and verify keys.  PGP developers are moving toward systems that use automated key distribution and authentication, with the traditional manual trust decisions left to a small minority with specialized needs.

\paragraph{Incremental improvement is still possible.}

Most email users trust their mailbox providers with plaintext email. While link encryption and domain authentication are available, vulnerabilities to active attacks and a lack of adoption leave email in transit subject to eavesdropping and message forgery. Providers could create an interoperable hosted S/MIME standard to automate provider-to-provider confidentiality and integrity, while still working within the threat model of a trusted mailbox provider. Unlike end-to-end encryption, server-based search, content-filtering, and persistent/portable mailbox access would be supported. Easy-to-deploy tools are needed to ensure the solution is not a barrier to entry for small providers.

\paragraph{Secure messaging is only a partial answer.} Messaging protocols are walled gardens, allowing proprietary protocols that are interactive and supported by central servers. This enables automated encryption for users, including automatic key exchange via a trusted key server and automatic end-to-end encryption of messages~\cite{unger2015sok}. Using a trusted key server means that users may be unaware of the security and usability tradeoffs they are making.  Users of secure messaging applications are typically only warned to check the encryption keys if they change, and numerous studies have shown that these applications fail to help users understand how to do this successfully~\cite{abu2017security,schroder2016signal,vaziripour2017you}. Security experts recommend encrypting all messages, however some applications make encryption optional, resulting in many users failing to turn encryption on~\cite{vaziripour2018survey}.

Further, email's open nature gives it fundamentally different uses than messaging, including easily communicating with strangers, sending long, content-rich messages, permanently archiving messages, searching past conversations, and attaching files. While email's additional features are part of the reason ubiquitous end-to-end encryption is so elusive, they are also why email is likely to continue to be a primary form of communication on the Internet for years to come.

%\paragraph{Choice of identity provider leads to key discovery problems.}

% JC: Most other discussion points are broad statements 

%Open systems provide the ability for users to choose their identity provider, but generally struggle with effortless encryption key discovery. An illustration of this tradeoff can be seen by examining Enhanced Certificate Transparency. The authors of Enhanced CT suggest that a user's email provider can naturally serve as their identity provider. This makes encryption key discovery easy because any email client can parse a recipient's email address and translate the domain name into the identity provider, \eg using DNS. The authors also mention that users could choose their identity provider, but do not design a critical piece of the system---mapping a user's email address to their preferred identity provider at that point in time.  This is a non-trivial problem. Autocrypt and Key Continuity use TOFU key exchanges for this reason---it's a simple way to allow for user choice of identity provider, with easy key discovery, though this sacrifices the ability to easily validate and thus trust keys. It remains to be seen if this problem can be solved in a way that a system would receive full marks for all usability properties

\paragraph{Vulnerable users are not well served.}

Aside from vulnerable users, every stakeholder represents a class of user that has their needs met by at least one system available today. Typical users are served by current offerings from email service providers. Enterprises (and their employees) are served by corporate S/MIME, which provides a combination of security, utility, and usability that matches their priorities. Deployment cost are likely what hinders its broader adoption among enterprises.  Privacy enthusiasts are served by secure webmail services, with their stronger emphasis on end-to-end encryption and good usability, while sacrificing utility to meet these priorities. In contrast, there is no system that clearly serves vulnerable users well. PGP is perhaps the best option, given its use by investigative journalists~\cite{romera2018icij}, but it does not meet all the security priorities of vulnerable users. No system except for remailers provides sender pseudonymity, and these do not typically meet other security properties important to vulnerable users. The small size and desire for anonymity among members of this  stakeholder group (journalists, dissidents, whistleblowers, survivors of violence, informants, under-cover agents, and even criminals) does not lend itself to commercial solutions, and volunteer organizations in this area have historically struggled.

\section{Research and Development Directions}

Improving the security of email is important to us. In this section, we briefly outline several avenues for future research and development.

\paragraph{Interoperability.}

Interoperability among secure email systems is a complex topic. Email evolved into an open system decades ago, allowing anybody to email anyone else. Thus, a justifiable user expectation is that secure email should likewise be open. However, we are far from achieving this today with secure mailbox providers (recall Section~\ref{sec:webmail}), since the primary secure systems in use are walled gardens, as either online services and/or dedicated software clients. Using standardized cryptographic suites is a small step but systems should also allow key (and key server) discovery between services (\eg ProtonMail-esque mailboxes to enterprise S/MIME certificate directories).

Interoperability introduces challenging issues around privacy, spam, and trust. Enterprises and providers are unwilling to expose the public keys of their users to outside queries. Encrypted spam, and other kinds of malicious email, can evade standard content filtering techniques that work on plaintext. Different systems operate under different trust models. While the web has built a system based on global trust, this requires only one-way trust of the web server, whereas secure email involves two-way trust between individuals and organizations. Simply adopting the web's CA trust model would be unlikely to yield a workable system, given the challenges that remain still largely unsolved with this model~\cite{CO13}. Technically a system based on a CA alternative (\eg trust-on-first-use) could interoperate with a different system (\eg  certificate directory) but typical users are unlikely to comprehend the difference in trust even if communicated to them, and the entire system could end up with weakest link security. Even if formats and protocols were universally agreed upon, it is not clear whether interoperability is always desired or meaningful. Finally, opening any system to interoperability means users will need help deciding which organizations or providers to trust to provide correct public keys. We argue it is both infeasible and unnecessary to expect that every individual or organization can be globally trusted by the others.

We advise future work on a much more limited goal of establishing trust among communicating parties when they need it. Any individual user or organization has a relatively small set of other users or organizations that it needs to trust. Developing infrastructure and protocols with this end in mind would appear to be necessary to leverage any gains made in technical interoperability.

\paragraph{Content inspection on encrypted email.}

Another major problem for secure email is coping with spam and malware.  Even if interoperability was a solved problem, authentication of an email sender is not the same as authorization to send email~\cite{bellovin2004look}, and building a system that provides the former but not the latter simply means users will get authenticated spam and phishing emails. End-to-end encryption systems without sufficient spam prevention for users are impractical, since both email providers and users lack an incentive to use such a system.

One possibility is to try to work around this problem. A secure email client could accept encrypted email only from regular or accepted contacts; rejecting encrypted email from unapproved senders could serve as a viable substitute for spam and malware filtering. Spam and malware could still be propagated by compromising accounts and spreading it to others who have approved those users, but the attack surface would be significantly limited. However, email providers are not likely to embrace such a system since it arguably offers less spam and malware protection for users than current practice.

A better alternative might be to build secure email systems that allow for server-side content processing even when private keys are only accessible to users. One possibility is to develop improved methods for processing on data that is encrypted~\cite{song2000practical,gentry2009fully,Kam15}. Alternatively, clients could send encrypted email and a decryption key to a trusted cloud computing environment~\cite{santos2009towards,pasquier2017camflow}, perhaps based on trusted execution platforms where the email could be decrypted and filtered for malware and spam. Likewise, a trusted computing environment could be used for storing and searching archives. Another possibility is to move email storage to edge devices owned by an end-user where content processing can be performed, with encrypted backup in the cloud to provide fault tolerance and portability.

 \paragraph{Auditing identity providers.}

Providing an auditable certificate directory or key server enables a system to provide a public key audit trail, responsive public key revocation, and effortless public key verification. However, additional work is needed to ensure such a system can meet its goals. For example, consider auditing systems like Certificate Transparency and CONIKS~\cite{ryan2014enhanced,melara2015coniks,basin2015arpki}. When it is a user's personal public key that is audited in such a system, the system must also then provide a usable method for users to monitor the public keys being advertised. In the case that a client's system notices that an unauthorized key is advertised for them, the system needs a method for the user to whistleblow and have the offending key revoked. Additionally, if the user's own identity provider has equivocated, then the user needs a method for being informed of this in a trustworthy manner and then being guided on choosing a new identity provider. If the identity provider is also their email provider, then they will also need to choose a new email provider. These auditing systems are promising and would benefit from further development and study to the point where we can be confident that it will be easy for users to accomplish these tasks.

\paragraph{Increasing trust.}

Recent work has shown that even with the proliferation of secure messaging applications, there is still a gap in how users perceive the effectiveness of security technology~\cite{abu2017obstacles,dechand2019encryption}. Users overestimate the capabilities of attackers and underestimate the strength of encryption technology, resulting in a lack of trust in applications that claim to protect their privacy. It is debatable whether this lack of trust is misplaced---the best cryptography cannot protect against errors in implementations or breaches that expose data that is stored unencrypted. Users have a healthy skepticism of general software and technology when they pay attention to highly publicized security failures. This is further complicated by `snake-oil' security and encryption tools that do not offer concrete benefits. Nevertheless, users are better off using encryption if they are going to communicate sensitive data online. Thus, user lack of trust in encryption is a major obstacle to overcome.

Trust is a longstanding challenge in computing~\cite{cra2003}. Secure messaging is only secure if you trust WhatsApp, for example, to exchange keys properly, or if you know enough to verify exchanged keys manually, or if you trust your messaging partners not to reveal the content of your messages. Yet the biggest success to date in getting users to adopt secure communication---the use of secure messaging applications---is not due to users choosing security or privacy but because users migrate to applications with large user bases and convenient functionality, which happen to use end-to-end encryption~\cite{abu2017obstacles}. It is not clear how email can follow the same path. Getting users to adopt secure email services may require gains in user understanding of risks and trust in solutions that mitigate those risks. The field of risk communication which has been used successfully for many years in public health, may offer a path toward helping users understand and cope with online security risks~\cite{nurse2011trustworthy,wu2019something}.

\paragraph{Removing private key management barriers.}

%Our analysis of key management schemes focused on public key distribution and revocation, mirroring the emphasis of work in the academic and developer communities.  However, t

There are numerous open questions regarding how typical (non-enterprise) users~\cite{ruoti2019johnny} will manage the full key life cycle, which includes private key storage, expiration, backup, and recovery~\cite[\S13.7]{menezes1996handbook}. These questions are complicated by issues such as whether to use separate keys for encrypting email during transmission, as opposed to those for long-term storage~\cite{chandramouli2016trustworthy}.
%, how much automation can be used to simplify key management, and how much users need to know in order to safeguard keys.
Storing keys in trusted hardware where they cannot be exfiltrated solves some storage issues, but also requires users to create backup hardware keys and revoke keys stored in lost or stolen devices. It is worth noting that major browsers and operating systems now support synchronizing passwords across user devices (under a user account with the provider), and one part of solving key management problems may involve using similar techniques to synchronize private keys.

% JC: I'm not crazy about below. If E2E encrypted, a lost key can be replaced but you still lose access to all old email, as well as new email before your new key is advertised. So it is sort of "account recovery" but not what people generally think (I log in after recovery and everything is the way I left it):

% Another option is to extend S/MIME-based systems primarily used by enterprises so that they are also available to users. A primary advantage of this approach is that if a user loses their private key (\eg by losing their device storing the key) they are not locked out of their accounts---they can simply get a new certificate issued by recertifying their identity with a Certification Authority. This requires trusting the authority, but in return users could receive help with both account recovery and revocation. For this to work, users need more usable methods to interact with Certification Authorities than are available now.

\paragraph{Addressing archive vulnerability.}

One of the consequences of high-profile phishing attacks in recent years has been the digital theft of the extensive information stored in long-term email archives of various individuals, companies, and organizations. It is ironic that the most active areas of research into securing email are largely orthogonal to the email security issues reported in the news. While data leaks might be categorized as a general data security issue, the way email products and architectures are designed (\eg emails archived by default, mail servers accessible by password) are inculpatory factors. Research on technical solutions, revised social norms about email retention, and other approaches could be helpful in addressing this issue.

% !TEX root = ../main.tex

\section{Concluding Remarks}

Deployment and adoption of end-to-end encrypted email continue to face many technical challenges, particularly related to key management. Our analysis
indicates that conflicting interests among stakeholders explains the fragmented nature of existing secure email solutions and the lack of widespread adoption. This suggests it is time to acknowledge that a one-size-fits-all (\ie for all target scenarios, environments, and user classes) solution or architecture will not emerge. In particular, we find the strongest conflicts among stakeholders over exceptional access, sender pseudonymity, server-side content-processing, and persistent access (\ref{col:persistent-access}). In each case, at least one stakeholder strongly prioritizes one of these properties while another strongly opposes it.

In this light, a significant barrier to progress is opposition to any new product or service that does not meet one stakeholder's particular needs, though it works well for others. A path forward is to acknowledge the need for alternate approaches and support advancement of alternatives in parallel. Divided communities and differing visions can lead to paralysis if we insist on a single solution, but it can also be a strength if we agree that multiple solutions can co-exist.

% Phishing remains a major problem~\cite{khonji2013phishing}, particularly spearphishing attacks against high-value targets, as evidenced by high-profile breakins.

% = = = = = Acknowledgments = = = = = %
\section*{Acknowledgments}
J. Clark acknowledges funding from the NSERC/Raymond Chabot Grant Thornton/Catallaxy Industrial Research Chair and his Discovery Grant. P.C. van Oorschot acknowledges NSERC funding for both his Canada Research Chair and a Discovery Grant. K. Seamons and D. Zappala acknowledge support by the National Science Foundation Grant No. CNS-1816929.

% = = = = = Bibliography = = = = = %

\bibliographystyle{ACM-Reference-Format}
\bibliography{bib/main.bib}

% = = = = = End Notes = = = = = %

\appendix
% !TEX root = ../main.tex

\section{Cryptographic Primitive Ratings Explanation}
\label{appx:crypto}

In this section we specify why each cryptographic primitive received the ratings it did in Table~\ref{tab:taxonomy-crypto}.

\subparagraph{\ref{cp:plaintext} \textit{\ref{cp-text:plaintext}}}
Plaintext email provides no protection from either eavesdropping (\ref{col:confidentiality}) or message tampering/injection (\ref{col:integrity}). Similarly, it does not provide pseudonymity (\ref{col:privacy}). As email is available in plaintext to the email provider, it allows for server-side content processing (\ref{col:content-processing}). Because it is already deployed it requires no changes to clients (\ref{col:email-client}) or servers (\ref{col:email-server}).

\subparagraph{\ref{cp:link-enc} \textit{\ref{cp-text:link-enc}}}
Link encryption email provides limited protection from eavesdropping (\ref{col:confidentiality}) and message tampering/injection (\ref{col:integrity}), earning a rating of no support because all intermediate mail transfer agents (MTAs)---not just mail providers---can read and/or modify the messages they receive. It does not provide pseudonymity (\ref{col:privacy}). Email is available in plaintext at the email provider, supporting server-side content processing (\ref{col:content-processing}). To our knowledge, all competent modern clients implement link encryption for communication between users and their mail providers (\ref{col:email-client}). While many email servers support link encryption, it is not enforced~\cite{durumeric2015neither,holz2016tls} by default so we rate it as not supporting the no email server updates property (\ref{col:email-server}).

\subparagraph{\ref{cp:mta-enc} \textit{\ref{cp-text:mta-enc}} and \ref{cp:mta-sig} \textit{\ref{cp-text:mta-sig}}}
Provider-to-provider encryption provides protection against eavesdropping (\ref{col:confidentiality}), whereas provider-to-provider signatures protect against message tampering/injection (\ref{col:integrity}). Because the sender's and recipients' mail providers can still see and/or modify the body of email messages, only partial support is awarded. These schemes do not provide pseudonymity (\ref{col:privacy}). Email is available in plaintext to the email provider, supporting server-side content processing (\ref{col:content-processing}). Because security operations occur at the mail providers, this primitive requires changes to the email servers (\ref{col:email-server}) but not email clients (\ref{col:email-client}).

\subparagraph{\ref{cp:layered} \textit{\ref{cp-text:layered}}}
Layered encryption provides protection from eavesdropping (\ref{col:confidentiality}), except at the recipients' email provider where the plaintext email is finally available. It provides limited protection against message tampering but not injection (\ref{col:integrity}), thus earning a rating of no support. This schemes does provide protection against mail servers linking senders or recipients and can be used to encrypt even routing portions of the envelope, receiving a rating of full support for pseudonymity (\ref{col:privacy}). As text is not available in plaintext to the sender's email provider, server-side content processing is not possible for the sender, though it may be possible for the recipients' email providers (\ref{col:content-processing}). This scheme requires changes both at the client (i.e., sender) (\ref{col:email-client}) and mail servers (\ref{col:email-server}).

\subparagraph{\ref{cp:e2e-enc} \textit{\ref{cp-text:e2e-enc}} and \ref{cp:e2e-sig} \textit{\ref{cp-text:e2e-sig}}}
End-to-end encryption provides full protection against eavesdropping (\ref{col:confidentiality}), whereas end-to-end signatures provide full protection against message tampering/injection (\ref{col:integrity}). These schemes do not provide pseudonymity (\ref{col:privacy}), as much of the meta-data (e.g., subject line) in both the headers and the envelope is neither encrypted or signed. Since email is already encrypted by the time it reaches the email provider, server-side content processing is unavailable (\ref{col:content-processing}). As operations occur at the client software, it requires changes to the email clients (\ref{col:email-client}) but not email servers (\ref{col:email-server}).
% !TEX root = ../main.tex

\section{Key Management Scheme Ratings Explanation}
\label{appx:keymanagement}

In this section we specify why each key management scheme received the ratings it did in Table~\ref{tab:taxonomy-key-management}.

\subparagraph{\ref{km:ca} \textit{\ref{km-text:ca}}}
Certification Authorities (CAs) require that users (with the help of client-side software) manage their own private keys (\ref{col:private-key}) and as such there is no backup of those private keys by the CA (\ref{col:persistent-access}).
CAs do not provide public key audit trails (\ref{col:public-key-audit}), but do publish revocation information for certificates (\ref{col:revocation}).
As CAs already exist, there is no need for new non-email infrastructure (\ref{col:infrastructure}).
To send encrypted email, a client must somehow obtain the certificate for the
recipient, containing their public key. Certificates could be sent with each outgoing email, but since this requires an extra step this system receives partial support for
effortless key discovery (\ref{col:key-discovery}). Once obtained, the authenticity of a public key is automatically validated (\ref{col:key-validation}) using the Certification Authority system and the trusted root store. Checking the revocation status of a public key can also be automated, so revocation does not affect our ratings.

\subparagraph{\ref{km:ca-directory} \textit{\ref{km-text:ca-directory}}}
A certificate directory extends \ref{km:ca}, adding full support for effortless key discovery (\ref{col:key-discovery}).
% This also lowers the bar for initial communication (\ref{col:enrollment}), though it still requires that recipients generate a key pair before they can receive encrypted email.
Certificate directories are additional non-email infrastructure that require deployment (\ref{col:infrastructure}).

% \subparagraph{\ref{km:ca-escrow} \textit{\ref{km-text:ca-escrow}}}
% A certificate directory with private key escrow extends \ref{km:ca}, adding full backup of private keys (\ref{col:persistent-access}). % and also allowing for simple initial communication (\ref{col:enrollment}) through the use of auto-generated key pairs for users who haven't yet adopted the scheme.
% The certificate directory and key escrow servers are additional non-email infrastructure that require deployment (\ref{col:infrastructure}).

\subparagraph{\ref{km:mkd} \textit{\ref{km-text:mkd}}}
Private keys are only accessible to users ({\ref{col:private-key}}), but the lack of any new infrastructure (\ref{col:infrastructure}) prevents a public key audit trail ({\ref{col:public-key-audit}}), responsive revocation ({\ref{col:revocation}}), or private key recovery (\ref{col:persistent-access}).
Persistent access to email is not supported since the loss of the user's private key renders the encrypted messages unreadable ({\ref{col:persistent-access}}).
Because users must manually share and verify public keys, a rating of no support is given for each usability property (\ref{col:key-discovery}, \ref{col:key-validation}).

\subparagraph{\ref{km:mkd-web} \textit{\ref{km-text:mkd-web}}}
This scheme extends \ref{km:mkd}, adding trusted introducers making it possible to validate keys without meeting in person as long as users share a common set of trusted introducers.
We do not believe that most users will be able to correctly configure mutual sets of trusted introducers, leading to many users being unable to validate each other's keys. As such we rate this scheme as having no support for effortless key validation ({\ref{col:key-validation}}).

\subparagraph{\ref{km:mkd-server} \textit{\ref{km-text:mkd-server}}}
This scheme extends \ref{km:mkd-web} by providing a central key server that allows for effortless public key discovery ({\ref{col:key-discovery}}).
It does not address the issues with trusted introducers (see \ref{km:mkd-web}), and so continues to be rated as having no support for effortless public key validation (\ref{col:key-validation}).
The key packet server is additional non-email infrastructure requiring deployment (\ref{col:infrastructure}).

\subparagraph{\ref{km:trusted-server} \textit{\ref{km-text:trusted-server}}}
Users manage the generation of their key pairs and are responsible for storing their private keys (\ref{col:private-key}), and as such backup of private keys is not provided by the system itself (\ref{col:persistent-access}).
The trusted public key server does not provide a public key audit trail (\ref{col:public-key-audit}), but does disseminate revocation information (\ref{col:revocation}).
The trusted key server is additional non-email infrastructure that requires deployment (\ref{col:infrastructure}).
While a trusted public key server does make the discovery of public keys effortless (\ref{col:key-discovery}), it receives only partial support for effortless validation (\ref{col:key-validation}) due to the trust required.

\subparagraph{\ref{km:trusted-ledger} \textit{\ref{km-text:trusted-ledger}}}
This scheme extends \ref{km:ca-directory} or \ref{km:trusted-server} by adding a public key audit trail (\ref{col:public-key-audit}).

\subparagraph{\ref{km:tofu} \textit{\ref{km-text:tofu}}}
Users private key are only accessible to the user ({\ref{col:private-key}}), but the lack of any new infrastructure (\ref{col:infrastructure}) prevents a public key audit trail ({\ref{col:public-key-audit}}), responsive revocation ({\ref{col:revocation}}), or private key recovery (\ref{col:persistent-access}).
Persistent access to email is not supported since the loss of the user's private key renders the encrypted messages unreadable ({\ref{col:persistent-access}}).
Because users must first receive email from someone, with an attached public key, before they can email them securely, this system receives partial support for effortless key discovery (\ref{col:key-discovery}).
It receives no support for effortless key validation because the user must manually verify keys if they wish to ascertain whether the key truly belongs to the purported owner \ref{col:key-validation}).

\subparagraph{\ref{km:shared-secrets} \textit{\ref{km-text:shared-secrets}}}
Because the encryption key is accessible to both the sender and the recipient,  the scheme has a rating of no support for ensuring the private key is accessible only to the sender ({\ref{col:private-key}}).
There are no public keys to audit ({\ref{col:public-key-audit}}) or revoke (\ref{col:revocation}).
Because there might be a need to revoke a compromised password, but no responsive way to do so (especially if the same secret is used for many messages) we award a rating of no support for revocation.
If the shared secret is lost, it might be possible to contact the other party to recover it (\ref{col:persistent-access}), but because this is not guaranteed to work, we award a rating of no support for persistent access.
This scheme requires no new non-email infrastructure (\ref{col:infrastructure}).
% We award a rating of full support for simple initial communication because the sender can encrypt a message for the recipient using a secret the sender generates without any action by the recipient ({\ref{col:enrollment}}).
The shared secret must be manually shared between the sender and recipient,
earning no support for effortless key discovery (\ref{col:key-discovery}).
In this scheme there are no true signatures or signing keys (\ref{col:key-validation}).

\subparagraph{\ref{km:ibe} \textit{\ref{km-text:ibe}}}
An IBE key server can generate the private key for any user (\ref{col:private-key}), allowing users to recover their private keys at will (\ref{col:persistent-access}).
The IBE key server provides no public key audit trail (\ref{col:public-key-audit}) and revocation is not easily supported (\ref{col:revocation})~\cite{boldyreva2008identity}.
The IBE trusted key server is additional non-email infrastructure that requires deployment (\ref{col:infrastructure}).
%Because public keys are mathematically derived from the email address and public parameters it is possible for a user to receive encrypted email before adopting the scheme (\ref{col:enrollment}).
The use of mathematically derived public keys allows for effortless public key discovery (\ref{col:key-discovery}) and validation (\ref{col:key-validation}).

\subparagraph{\ref{km:escrow} \textit{\ref{km-text:escrow}}}
Key escrow adds full support for persistent access (\ref{col:persistent-access}) to any scheme, while removing the property that private keys (or symmetric encryption keys) are available only to the user (\ref{col:private-key}).
The key escrow servers are additional non-email infrastructure that require deployment (\ref{col:infrastructure}).

% !TEX root = ../main.tex

\section{Secure Email Design Ratings Explanation}
\label{appx:systems}

In this section we specify why each secure email design scheme received the ratings it did in Table~\ref{tab:taxonomy-systems}.

\subparagraph{\ref{sys:plaintext} \textit{\ref{sys-text:plaintext}}}
This design provides no security but has full support on relevant utility ad deployability properties. This system design and other MTA-based approaches do not offer end-to-end encryption so they are not evaluated on properties related to key management or identity providers.

\subparagraph{\ref{sys:modern-email} \textit{\ref{sys-text:modern-email}}}
While links are encrypted, this does not prevent intermediary MTAs from reading the contents of a message, leading to this scheme being rated as having no support for confidentiality (\ref{col:confidentiality}).
The use of DKIM is rated as partial support for integrity (\ref{col:integrity}).
Many servers do not enforce TLS or use DKIM~\cite{durumeric2015neither,foster2015security,holz2016tls}, so this system design requires changes to email servers (\ref{col:email-server}).
DKIM requires the adoption of DNSSEC, resulting in a rating of no support for no new non-email infrastructure (\ref{col:infrastructure}).

\subparagraph{\ref{sys:mixminion} \textit{\ref{sys-text:mixminion}}}
This system design provides pseudonymity (\ref{col:privacy}), but since the recipients' mail providers can see the contents of the message, it provides only partial support for confidentiality (\ref{col:confidentiality}).
This system design requires custom clients (\ref{col:email-client}) and new mixminion servers (\ref{col:infrastructure}), leading to a rating of no support for both properties.

\subparagraph{\ref{sys:smime} \textit{\ref{sys-text:smime}}}
While messages are encrypted end-to-end, the private key (\ref{col:private-key}) is accessible to the enterprise, resulting in partial support for confidentiality (\ref{col:confidentiality}) and (\ref{col:integrity}), as well as a rating of no support for preventing exceptional access (\ref{col:exceptional-access}).
Even though ownership of the private key would allow server-side content processing (\ref{col:content-processing}), this is not done in practice.
Escrow of the private key allows users to retrieve lost keys, providing persistent access to their email (\ref{col:persistent-access}).
Sender and recipient email addresses are unencrypted and do not provide pseudonymity (\ref{col:privacy}).
Currently, corporate S/MIME solutions do not support public key audit trails (\ref{col:public-key-audit}), but do provide responsive revocation (\ref{col:revocation}).
This system design does nothing to make it easy to detect phishing (\ref{col:phishing}).

Users do not have choice of email provider (\ref{col:provider-choice}) or Certification
Authority (\ref{col:server-choice}).
While many clients already support this system design (e.g., Outlook, Apple Mail), support is not universal, and so we rate it as needing client software updates  (\ref{col:email-client}).
There are no changes needed for email servers (\ref{col:email-server}), but it is necessary to deploy servers that support key escrow (\ref{col:infrastructure}).
Users of this system design can effortlessly discover (\ref{col:key-discovery}) public keys for users in the same organization, but validation (\ref{col:key-validation}) requires trusting the Certification Authority.

\subparagraph{\ref{sys:hosted-smime} \textit{\ref{sys-text:hosted-smime}}}
%Public key auditability (\ref{col:public-key-audit}) is provided through the use of Certificate Transparency.
Because email is encrypted and decrypted by the email server, it is possible for that server to perform content processing on that material (\ref{col:content-processing}).
Because it is the server, not the client, that handles encryption, it is also the server (\ref{col:email-server}), not the client (\ref{col:email-client}), that needs to be modified to make this system design work. Because a user needs to receive an email first in order to discover the encryption key for the sender of that email, the system design receives partial support for effortless encryption key discovery (\ref{col:key-discovery}).

\subparagraph{\ref{sys:secure-webmail} \textit{\ref{sys-text:secure-webmail}}}
This system design uses end-to-end encryption, so it receives full support for protection from eavesdropping (\ref{col:confidentiality}) and tampering (\ref{col:integrity}).
Because the service stores the user's (password-protected) private key, we rate it as having only partial support for that property (\ref{col:private-key}) and for prevention of exceptional access (\ref{col:exceptional-access}), as well as partial support for persistent access to email (\ref{col:persistent-access}).
Because ProtonMail offers a new service, we rate them as having no support for no new infrastructure needed (\ref{col:email-server}).
% As this system design requires that recipients adopt not just a new email client, but an entirely new email provider there is not support for simple initial communication (\ref{col:enrollment}).
Using a centralized key server for their users provides full support for effortless public key discovery (\ref{col:key-discovery}), but only partial support for validation (\ref{col:key-validation}) since the server is trusted.

\subparagraph{\ref{sys:pgp} \textit{\ref{sys-text:pgp}}}
This system design uses end-to-end encryption, so it receives full support for protection from eavesdropping (\ref{col:confidentiality}) and tampering (\ref{col:integrity}).
Private keys are accessible only to their respective users (\ref{col:private-key}), so it receives full support for preventing exceptional access (\ref{col:exceptional-access}), but effectively prevents server-side content processing (\ref{col:content-processing}) and persistent access to email in the case of private key loss (\ref{col:persistent-access}).
Sender and recipient email addresses are unencrypted and do not provide pseudonymity (\ref{col:privacy}).
Key packet servers do not currently provide a public key audit trail (\ref{col:public-key-audit}), nor do they currently provide effective and responsive revocation (\ref{col:revocation}).

This system design requires that users install new software (\ref{col:email-client}) and that key packet servers (\ref{col:infrastructure}) are deployed, but does not require changes to email servers (\ref{col:email-server}).
This system design allows users their choice of email provider (\ref{col:provider-choice}) and key server (\ref{col:server-choice}).
Users of this system design can't effortlessly discover (\ref{col:key-discovery}) public keys for users since they must know which key packet server the recipient uses.
Likewise, the difficulty of setting up trusted introducers leads to a rating of no support for effortless validation (\ref{col:key-validation}).

\subparagraph{\ref{sys:tofu} \textit{\ref{sys-text:tofu}}}
Because users must email each other to share keys, but otherwise discovery is automated, this system design receives partial support
for effortless public key discovery (\ref{col:key-discovery}). Because keys are not validated without some additional manual comparison it receives no support for
effortless key validation (\ref{col:key-validation}).
%While using email as a channel to share public keys is a usability improvement over in-person sharing, it does not ensure that a sender will be able to obtain an encryption key for the recipient they wish to communicate with.
%While this scheme does not warrant a higher rating, we note that among closed groups of contacts it is possible that public keys will automatically be shared amongst the group before there is a desire to encrypt email.
%Lastly, the use of TOFU key binding means that validating public keys (\ref{col:key-validation}) is effortless.

\subparagraph{\ref{sys:johnny2} \textit{\ref{sys-text:johnny2}}}
Ratings for key continuity are identical to Autocrypt, except that the system design provides partial support for detecting phishing (\ref{col:phishing}) since key continuity
provides a warning when a different certificate is seen.

\subparagraph{\ref{sys:ect} \textit{\ref{sys-text:ect}}}
This system design uses end-to-end encryption, so it receives full support for protection from eavesdropping (\ref{col:confidentiality}) and tampering (\ref{col:integrity}).
Private keys are accessible only to their respective users (\ref{col:private-key}), so it receives full support for preventing exceptional access (\ref{col:exceptional-access}), but prevents server-side content processing (\ref{col:content-processing}) and persistent access to email in the case of private key loss (\ref{col:persistent-access}).
The use of a Certification Authority provides responsive revocation (\ref{col:revocation}), and
the use of Certificate Transparency provides an audit trail for the user's public key (\ref{col:public-key-audit}). Key discovery is effortless (\ref{col:key-discovery}) because the email provider for a recipient is assumed to be the identity provider as well. Key validation is likewise effortless (\ref{col:key-validation}) because of the CA system design and Certificate Transparency.

\subparagraph{\ref{sys:virtru} \textit{\ref{sys-text:virtru}}}
Because the email server does not have access to  users' private keys, and the key escrow server does not have access to users' messages, this scheme is rated as having full support for confidentiality (\ref{col:confidentiality}) and integrity (\ref{col:integrity}).  However, both parties could be coerced into revealing the contents of a user's message, so it is rated as having partial support for exceptional access (\ref{col:exceptional-access}). The use of
centralized, symmetric encryption provides responsive revocation because a compromised encryption key can be easily deactivated (\ref{col:revocation}), and escrow provides persistent access to email (\ref{col:persistent-access}).
The system design makes it easy to discover the (symmetric) encryption key (\ref{col:key-discovery}), but validation (\ref{col:key-validation}) is rated as no support since there
is no mechanism to verify the authenticity of the encryption key, as could
be done in a public key system design.
%Several other properties are left blank to indicate they do not apply because the system design does not use public key cryptography.

%The centralized key escrow system design rates as having full support for effortless public key discovery (\ref{col:key-discovery}), but validation is rated as partial support due to use of a trusted key server (\ref{col:key-validation}).
%The key escrow system design can also be used to automatically generate public keys for users that haven't enrolled in the system design, allowing for full support of simple initial communication (\ref{col:enrollment}).

%\subparagraph{\ref{sys:pwm} \textit{\ref{sys-text:pwm}}}
%This system design is rated nearly identically to Virtru on all properties, since it also uses a centralized key server that retains access to the
%user's private key, with the exception of responsive revocation.

%\subparagraph{\ref{sys:confidante} \textit{\ref{sys-text:confidante}}}
%The key server in this system design provide both a public key audit trail (\ref{col:public-key-audit}) and responsive key revocation (\ref{col:revocation}).
% As all users trust the same key server, and not potentially disparate groups of trusted introducers, initial communication (\ref{col:enrollment}) is possible after the recipient has installed the system design, leading to a rating of partial support.
%The use of a single global key management servers provides full support for effortless key discovery (\ref{col:key-discovery}) and validation (\ref{col:key-validation}).

\end{document}